\documentclass[transmag]{IEEEtran}
\usepackage{latexsym}
\usepackage{graphicx}
\usepackage{amsfonts,amssymb,amsmath}
\usepackage{hyperref}
\usepackage{xcolor}
\usepackage{tikz}

\usepackage{textcomp}
\def\BibTeX{{\rm B\kern-.05em{\sc i\kern-.025em b}\kern-.08em
		T\kern-.1667em\lower.7ex\hbox{E}\kern-.125emX}}

\usepackage[noadjust]{cite}

\makeatother

\providecommand{\theoremname}{Theorem}

\newtheorem{lem}{\protect\lemmaname}

\makeatother

\providecommand{\lemmaname}{Lemma}

\newtheorem{prop}{\protect\propositionname}

\makeatother

\providecommand{\propositionname}{Proposition}

\makeatother

\providecommand{\corname}{Corollary}

\newtheorem{rem}{\protect\remname}

\makeatother

\providecommand{\remname}{Remark}

\begin{document}
	
	\title{Reconfigurable Intelligent Surface Optimal Placement in Millimeter-Wave Networks}
	
	\author{Konstantinos Ntontin, \IEEEmembership{Member, IEEE}, {Alexandros--Apostolos A. Boulogeorgos}, \IEEEmembership{Senior Member, IEEE}, \\  Dimitrios Selimis, \IEEEmembership{Graduate Student Member, IEEE},  Fotis Lazarakis,   \\ 
		Angeliki Alexiou, \IEEEmembership{Member, IEEE}, and Symeon Chatzinotas, \IEEEmembership{Senior Member, IEEE}

		\thanks{K. Ntontin is with the Interdisciplinary Centre for Security, Reliability and Trust (SnT) – University of
			Luxembourg, L-1855 Luxembourg and the  Wireless
			Communications Laboratory of the Institute of Informatics and Telecommunications, National Centre for Scientific Research–``Demokritos,'' Athens, Greece. E-mail: kostantinos.ntontin@uni.lu.}
		
		\thanks{A.-A. A. Boulogeorgos and A. Alexiou  are with the Department of Digital Systems,
			University of Piraeus
			Piraeus 18534 Greece. E-mails:  al.boulogeorgos@ieee.org,  alexiou@unipi.gr.
		}
		
		\thanks{D. Selimis and F. Lazarakis are with the Wireless
			Communications Laboratory of the Institute of Informatics and Telecommunications, National Centre for Scientific Research–``Demokritos,'' Athens, Greece. E-mails: $\left\{\text{dselimis, flaz}\right\}$@iit.demokritos.gr.}
		
		\thanks{S. Chatzinotas is with the Interdisciplinary Centre for Security, Reliability and Trust (SnT) – University of
			Luxembourg, L-1855 Luxembourg. E-mail: symeon.chatzinotas@uni.lu.}
		\thanks{This work was supported by the European Commission's Horizon 2020 research and innovation programme (ARIADNE) under grant agreement No. 871464 and the Luxembourg National Research Fund (FNR) under the CORE project RISOTTI.}
		\thanks{The associate editor coordinating the review process and accepting it for publication was Prof. Emil Bj\"{o}rnson.} 
	}

	\IEEEtitleabstractindextext{
		\begin{abstract}  
			This work discusses the optimal reconfigurable intelligent surface placement in highly-directional millimeter wave links. In particular, we present a novel system model that takes into account the relationship between the transmission beam footprint at the RIS plane and the RIS size. Subsequently, based on the model we derive the end-to-end expression of the received signal power and, furthermore, provide approximate closed-form expressions in the case that the RIS size is either much smaller or at least equal to the transmission beam footprint. Moreover, building upon the expressions, we derive the optimal RIS placement that maximizes the end-to-end signal-to-noise ratio. Finally, we substantiate the analytical findings by means of simulations, which reveal important trends regarding the optimal RIS placement according to the system parameters.  
		\end{abstract}

		
	}

	\maketitle
	
	\section*{Nomenclature}
	\addcontentsline{toc}{section}{Nomenclature}
	\begin{IEEEdescription}[\IEEEusemathlabelsep\IEEEsetlabelwidth{$V_1,V_2,V_3$}]
		\item[B5G] Beyond the Fifth Generation
		\item[FNBW] First-Null Beamwidth
		\item[HPBW] Half-Power Beamwidth
		\item[LoS] Line-of-Sight
		\item[mmWave] Millimeter Wave
		\item[NLoS] Non-Line-of-Sight
		\item[PIN] Positive-Intrinsic-Negative
		\item[RF] Radio-Frequency
		\item[RIS] Reconfigurable Intelligent Surface
		\item[RU] Reflection Unit
		\item[RX] Receiver
		\item[SNR] Signal-to-Noise-Ratio
		\item[TX] Transmitter 
	\end{IEEEdescription}
	
	\section{Introduction}\label{S:Intro}
	
	\subsection{Background and Related Works}
	
	Increasing data-rate demands have led current mobile-access networks relying on sub-6 GHz bands reach their limits in terms of available bandwidth. This bottleneck created the need to consider beyond-6 GHz bands for mobile-access networks. Currently, bands in the lower-end of the millimeter wave (mmWave) spectrum are used for point-to-point and point-to-multipoint line-of-sight (LoS) wireless backhaul/fronthaul and fixed-wireless access networks~\cite{ABI_Research}. Such deployments span the 30-100 GHz operational frequency range. However, the expected migration of future mobile-access networks to the 30-100 GHz band pushes the corresponding wireless backhaul/fronthaul links towards the beyond-$100$ GHz bands. Due to this, backhauling/fronthauling transceiver equipment vendors have performed LoS trials in the D-band ($130-174.8$ GHz), which showcase the potential of using it in such deployments~\cite{ETSI_D_Band}. Apart from LoS, street-level deployments in dense urban scenarios necessitate devising non-LoS (NLoS) solutions since LoS links may not always be available. However, despite the fact that according to measurements \cite{Rappaport_channel_capacity_mmwave}, \cite{Rappaport_mmWave_It_Will_Work!} NLoS communication through scattering and reflection from objects in the radio path is feasible in the 30-100 GHz range,  the higher propagation loss of beyond-100 GHz bands is likely to challenge this assumption.
	
	The conventional approach of counteracting NLoS links is by providing alternative LoS routes through relay nodes \cite{Laneman}. Although this is a well-established method to increase the coverage when the signal quality of the direct links is low, it is argued that it cannot constitute a viable approach for massive deployment, especially for mmWave networks. This is due to the increased power consumption of the active radio-frequency (RF) components in high frequencies that relays need to be equipped with \cite{Khawaja_mmWave_Passive_Reflectors}. Apart from relaying, communication through passive non-reconfigurable specular reflectors, such as
	dielectric mirrors, has been proposed as another alternative. Such an approach
	has the potential to be notably more cost efficient compared
	with relaying and has been documented at both mmWave and
	beyond-$100$ GHz bands~\cite{Khawaja_mmWave_Passive_Reflectors, THz_Passive_Reflectors}. Due to the highly dynamic nature of blockage at high
	frequencies together with the traffic conditions, which may necessitate fast rerouting of information within a network, it would be desirable that such reflectors can
	change the angle of departure of the waves so that they direct the beams towards different routes. However, passive reflectors are incapable of supporting the aforementioned functionality since the conventional Snell's law applies. Furthermore, even by enabling this functionality by means of mechanical steering of the passive reflectors, the resulting latency would substantially compromise the desired reliability. Based on the above, an intriguing
	question that arises is the following: \emph{Would it be possible
		to deploy reconfigurable reflectors that can arbitrarily steer the impinging beam based on dynamic blockage and traffic conditions and without compromizing the desired latency}?
	The answer is affirmative by considering the reconfigurable intelligent surface (RIS) paradigm.
	
	RISs are two-dimensional structures of dielectric material, which embed tunable reflection units (RUs) \cite{Wu_Zhang_1, Wu_Zhang_2, Ozdogan_Emil, Comparison_RIS_Relaying_Boulogeorgos, Kisseleff_RIS}. They constitute a substantially different technology than active relaying, due to the absence of bulky and power-hungry analog electronic components, such as power amplifiers. Additionally, their operation, in contrast with active relaying, does not require dividers and combiners, which can incur high insertion losses. By individually tuning the phase response of each individual RIS element, the reflected signals can constructively aggregate at a particular focal point, such as the receiver. Such a tuning can be enabled by electronic phase-switching components, such as positive-intrinsic-negative (PIN) diodes, RF-microelectromechanical systems, and varactor diodes, that are introduced between adjacent elements \cite{Basar_Reconfigurable_Intelligent_Surfaces}. Hence, RISs offer an alternative-to-relaying method for large-scale beamforming without the incorporation of high power consuming electronics and insertion losses involved by the additional circuitry. In practice, the RIS element phase shift can be controlled by a central controller through programmable software \cite{Basar_Reconfigurable_Intelligent_Surfaces}.
	
	Recognizing the unprecedented features that RISs can bring to beyond the fifth generation (B5G) wireless systems, a great amount of research effort has been put on analyzing, designing and optimizing RIS-aided wireless systems~\cite{Basar_Reconfigurable_Intelligent_Surfaces, Bariah2020,Wu2019, Jung2020, Zeng2020, Holographic_MIMO_RIS, RIS_assisted_multiuser_MISO, Marco_Di_Renzo_Survey_RISs}, as well as comparing them with their predecessors, i.e., relaying-aided ones~\cite{Comparison_RIS_Relaying_Boulogeorgos,Renzo2020,Huang_Reconf_Intell_Surfaces, Bjornson_Relaying}. In more detail,  in~\cite{Basar_Reconfigurable_Intelligent_Surfaces} and~\cite{Bariah2020} the authors introduced the idea of employing an RIS in order to mitigate the impact of blockage and steer the transmission beam towards the desired direction. Likewise, in~\cite{Wu2019} the authors presented the optimization framework for the maximization of the reception power in a RIS-aided system, assuming that all the RIS area can be used to reflect the induced electromagnetic wave. Moreover, in~\cite{Jung2020} the authors studied the asymptotic uplink ergodic capacity performance of an RIS-aided wireless system, while in~\cite{Zeng2020} the coverage of a downlink RIS-assisted network was studied, assuming that the entire RIS area can be used, and a strategy for maximizing the cell coverage by optimizing the RIS orientation and horizontal distance was proposed. In \cite{Holographic_MIMO_RIS}, the RIS empowered holographic multiple-input-multiple-output architecture is introduced, whereas in \cite{RIS_assisted_multiuser_MISO} the joint design of transmit beamforming matrix at the base station and the phase shift matrix at the RIS in a multiuser multiple-input-multiple-output setup is investigated. In addition, \cite{Marco_Di_Renzo_Survey_RISs} provides an extensive survey on RIS-related works in various domains. Finally, several works consider RISs operating as reflectors and show that sufficiently large RISs can outperform conventional active relays either is terms of rate or energy efficiency \cite{Comparison_RIS_Relaying_Boulogeorgos, Renzo2020, Huang_Reconf_Intell_Surfaces, Bjornson_Relaying}.

    \subsection{Motivation, Novelty, and Contribution}
	
	All the presented RIS-related works consider the case of the entire RIS area being illuminated by the transmitted beam. However, due to the highly directional transmissions in mmWave networks and the low manufacturing cost of RISs, which make them suitable, as it is envisioned, to cover a big portion of the facades of large structures, such as buildings, it is expected that in many cases only a part of the total RIS area is going to be illuminated. Based on this, our work is motivated by the need to answer the question of what the optimal RIS placement policy is, which can be seen as a network planning question, in the two cases of the RIS area being smaller and larger than the transmitted beam footprint. Summarizing, the technical contribution of the paper is as follows\footnote{This work constitutes an extension of \cite{Optimal_RIS_Placement_Eucap_2021}.}:
	\begin{itemize}
		\item We present a system model for RIS-aided highly directional mmWave links of fixed topology, such as  wireless backhaul/fronthaul links, and use electromagnetic theory to evaluate the received power in the general case of an RIS of arbitrary size. Of note, the presented analytical methodology can also find application in mobile mmWave networks, as it is elaborated in Section~\ref{S:Res}.
		
		\item Based on the resulting received-power expression, we provide approximate closed-form expressions for the cases in which the transmission footprint at the RIS plane is either much larger or smaller than the RIS.  According to the expressions, we evaluate the end-to-end SNR for both cases.
		
		\item We use the closed-form SNR expressions to analytically extract the policies for the optimum RIS placement that results in SNR maximization.
		 
		\item Finally, we provide an extensive simulation campaign in various scenarios in order to validate the analytical results and, furthermore, to provide design guidelines to the system designer from a practical point of view.
		
	\end{itemize}
	
	\emph{Organization}: The rest of this contribution is structured as follows: In Section~\ref{S:SM}, the system model is presented. In Section~\ref{S:SNR}, we firstly provide an expression for the transmission beam footprint at the RIS plane. Subsequently, based on it, through electromagnetic theory we compute the end-to-end received power. In addition, we provide closed-form approximate expressions for the received power and, correspondingly, for the end-to-end SNR, in the two cases of the RIS being either much smaller than then transmission beam footprint or larger. Finally, in the same section by leveraging the analytical SNR expressions we mathematically compute the optimal RIS placement that maximizes the SNR. Extensive numerical results that validate the analytical outcomes together with a discussion of how the system designer can use the presented results are provided in Section~\ref{S:Res}. Finally, Section~\ref{S:Conclusions} concludes this work by highlighting the most important findings and remarks.
	
	\emph{Notation}: For the convenience of the readers, recurrent parameters and symbols with their meaning are presented in Table~\ref{Recurrent_parameters_and_symbols}.
	
	\begin{table}[h]
		\label{Recurrent_parameters_and_symbols}
		\caption{Recurrent parameters and symbols.} 
		\centering 
		\scalebox{0.8}{
			\begin{tabular}{| c | c | c | } 
				\hline
				Parameter/Symbol& Meaning \\ [0.5ex]
				\hline
				\hline
				$f$& Carrier frequency  \\ [0.5ex]
				\hline
				$\lambda$& Wavelength \\ [0.5ex]
				\hline
				$P_{t}$& Transmit power \\ [0.5ex]
				\hline
				$W$& Signal bandwidth \\ [0.5ex]
				\hline
				$\mathcal{F}_{{\rm{dB}}}$& Noise figure\\[0.5ex]
				\hline
				$N_0$& Thermal noise power \\ [0.5ex]
				\hline
				$h_{s}$&RIS height with respect to the ground \\ [0.5ex]
				\hline
				$h_{t}$/$h_{r}$& TX/RX height with respect to the ground  \\ [0.5ex]
				\hline
				$D_{t}$/$D_{r}$ & TX/RX antenna diameter \\ [0.5ex]
				\hline
				$\phi_{0}$ & TX antenna FNBW \\ [0.5ex]
				\hline
				$\phi_{HPBW}$ & TX antenna HPBW \\ [0.5ex]
				\hline
				$e_{t}/e_{r}$ & TX/RX antenna aperture efficiency \\ [0.5ex]
				\hline
				$G_{t}^{max}/G_{r}^{max}$ & TX/RX antenna gain at the boresight \\ [0.5ex]
				\hline
				$G_{t,n}/G_{r,n}$ & TX/RX antenna gain with respect to the $n_{th}$ RU. \\ [0.5ex]
				\hline
				$\Gamma$ & Amplitude reflection coefficient of the RUs \\ [0.5ex]
				\hline
				$d_x$, $d_y$ & x-axis and y-axis length, respectively, of the RUs  \\ [0.5ex]
				\hline
				$\alpha$, $\beta$ & Radii of the TX beam elliptic footprint \\ [0.5ex]
				\hline
				$\epsilon$ & Eccentricity of TX beam elliptic footprint \\ [0.5ex]
				\hline
				$r_1$ & \shortstack{Distance between the center of the TX antenna and the center of \\ the TX footprint at the RIS plane}  \\ [0.5ex]
				\hline
				$r_2$ & \shortstack{Distance between the center of the TX footprint at the \\ RIS plane and the center of the RX antenna}  \\ [0.5ex]
				\hline
				$r_{1,n}$ & \shortstack{Distance between the center of the TX antenna and the $n_{th}$ RU}  \\ [0.5ex]
				\hline
				$r_{2,n}$ & \shortstack{Distance between the $n_{th}$ RU and the center of the RX antenna}  \\ [0.5ex]
				\hline
				$r_{1,h}$  & TX-RIS horizontal distance  \\ [0.5ex]
				\hline
				$r_{1,h}^{*}$  & Optimal RIS horizontal distance  \\ [0.5ex]
				\hline
				$r_h$  & TX-RX horizontal distance  \\ [0.5ex]
				\hline
				$\theta_i$, $\theta_r$ & \shortstack{Electromagnetic-wave incidence and departure angles, respectively,\\ with respect to the RIS center} \\ [0.5ex]
				\hline
				$\theta_{i,n}$, $\theta_{r,n}$ & \shortstack{Electromagnetic-wave incidence and departure angles, respectively,\\ with respect to the $n_{th}$ RU} \\ [0.5ex]
				\hline
				$S_s$ & RIS area \\ [0.5ex]
				\hline
				$S_{i}$ & Area of the TX beam elliptic footprint corresponding to the FNBW \\ [0.5ex]
				\hline
				$S_{HPBW}$ & Area of the TX beam elliptic footprint corresponding to the HPBW \\ [0.5ex]
				\hline
				$P_R$ & Received power \\ [0.5ex]
				\hline
				$\rho$ & SNR \\ [0.5ex]
				\hline
		\end{tabular}}
		\label{Recurrent_parameters_and_symbols} 
	\end{table}
	
	\section{System Model}\label{S:SM}

	\begin{figure}
		\centering
		{\includegraphics[width=3.4in,height=2.4in]{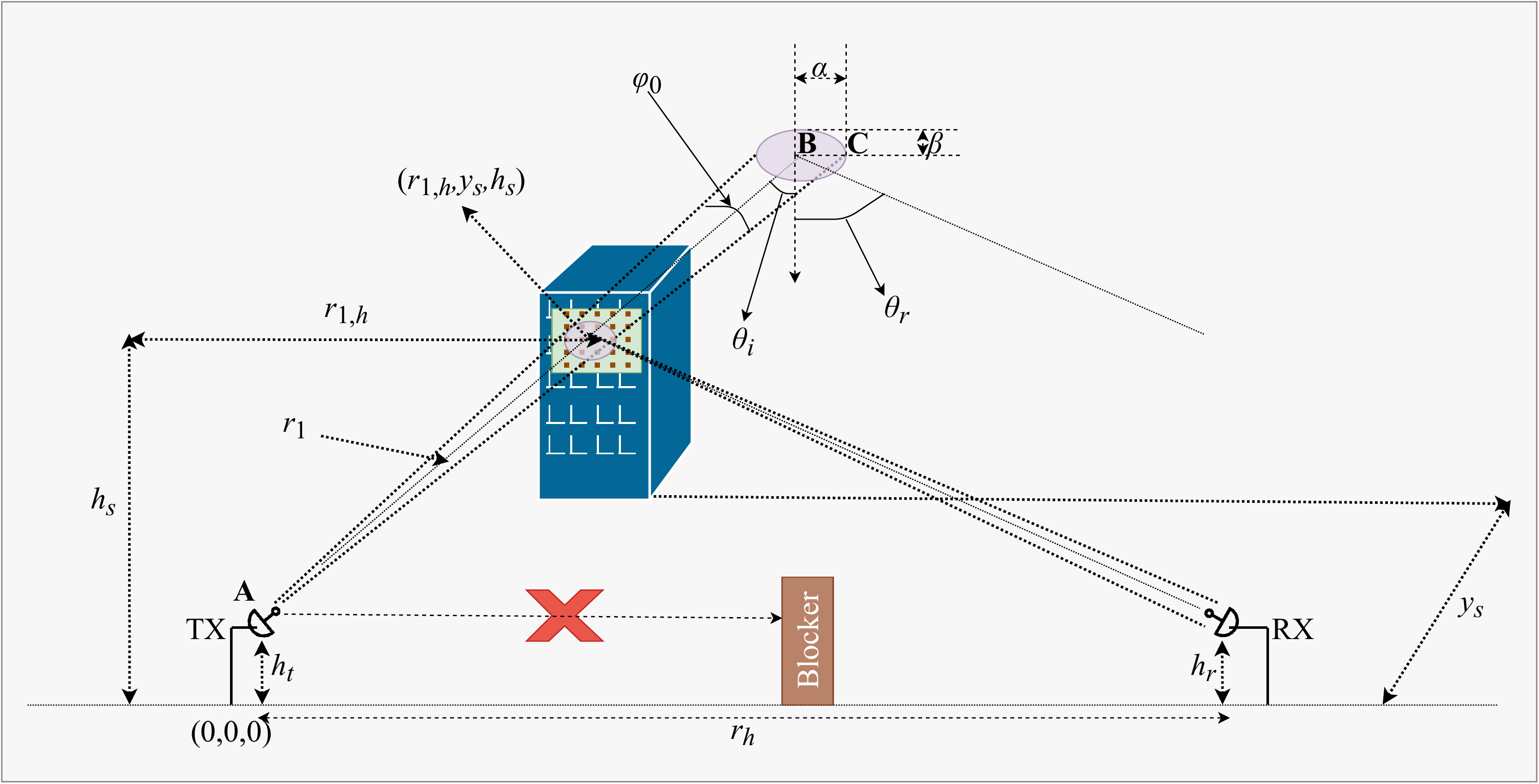}}
		\caption{System model.}
		\label{Fig:Communication_through_an_RIS}	
	\end{figure}
	
	As illustrated in Fig.~\ref{Fig:Communication_through_an_RIS}, we consider a fixed-topology street-level scenario, in which a TX communicates with a RX through an RIS. $r_{1,h}$, $r_{2,h}$, and $r_h$ are the horizontal TX-RIS, RIS-RX, and TX-RX distances, respectively, while $h_t$, $h_{s}$, and $h_{r}$ are the TX, RIS, and RX heights, respectively. $\theta_i$ and $\theta_r$ are the incidence and departure angles, respectively, of the electromagnetic wave with respect to the center of the illuminated area. The considered TX-RIS and RIS-RX blockage-free links are established in a mmWave band and constitute an alternative path to the direct TX-RX link that is assumed to be blocked. To countermeasure the high pathloss in this band, both the TX and RX are equipped with highly directional parabolic antennas with diameters $D_{t}$ and $D_{r}$, respectively. As a result, for $D_{t},D_{r}>>\lambda$, where $\lambda$ represents the wavelength,  their power radiation patterns $E_t\left(\phi\right)$ and $E_r\left(\phi\right)$, respectively, are given by \cite{anderson2003fixed}
	\begin{align}
	\label{electric_field_radiation_pattern}
	E_m\left(\phi\right)=\frac{2\lambda}{\pi D_m}\frac{J_1\left(\frac{\pi D_m \sin\left(\phi\right)}{\lambda}\right)}{\sin\left(\phi\right)}, \quad 0\le\phi<\pi/2
\end{align}for $m\in\{t, r\}$. $\phi$ is measured from the broadside direction ($\phi=0$) and $J_1\left(\cdot\right)$ is the first-order Bessel function of the first kind.
Hence, their gains, denoted by $G_t\left(\phi\right)$ and $G_r\left(\phi\right)$, respectively, are given by
\begin{align}
	G_m\left(\phi\right)&=\frac{e_m4\pi E_m^2\left(\phi\right)}{\int_0^{2\pi}\int_0^{\frac{\pi}{2}}E_m^2\left(\phi\right)\sin\left(\phi\right)d\phi d\theta}\\\nonumber&=4e_m\left(\frac{J_1\left(\frac{\pi D_m \sin\left(\phi\right)}{\lambda}\right)}{\sin\left(\phi\right)}\right)^2, \quad m\in\{t, r\},
\end{align}
where $e_{t}$ and $e_{r}$ denote the aperture efficiencies of the TX and RX antennas, respectively. Consequently, their maximum gain, denoted by $G_m^{max}$, that is obtained for $\phi=0$ is given by
\begin{align}
	G_m^{max}=e_m\left(\frac{\pi D_m}{\lambda}\right)^2, \quad m\in\{t, r\}.
\end{align}
Note that this type of antennas has been extensively used for wireless backhaul/fronthaul scenarios (see e.g.,~\cite{Backhaul_Fronthaul_Parabolic_Reflectors} and reference therein), due to their capability to support pencil-beamforming transmissions. Under such highly-directional transmissions, the three-dimensional antenna pattern can be modeled as a cone for half-power beamwidths (HPBWs), which we denote by $\phi_{HPBW}$, smaller than approximately 15$^{\circ}$~\cite[Ch. 12]{silver1984microwave}. 
	Furthermore, we assume that the TX and RX antennas can be mechanically steered, both in azimuth and elevation, towards the desired angle of transmission and reception, respectively, and they are pointing towards the center of the illuminated RIS region. 
	
	As far as the channel model is concerned, the assumption of mmWave links means that in the general case besides the direct LoS component several distinguishable multipath components also arrive at the RX either at the same or at different time instants depending on whether a narrowband or wideband model applies, respectively \cite{Hanzo_MmWave_Channel_Model}. However, when highly-directional antennas are employed at both the TX and RX sides, in the case of a wideband channel, for instance, that corresponds to common bandwidths at mmWave bands, there is virtually no delay spread according to real-world measurements \cite{Rappaport_MIllimeter_Wave_Measurements_and_Models}. Consequently, due to the considered fixed-topology scenario of this work with pencil-beam deployed antennas and the fact that the RISs are deployed in elevated positions with respect to the TX and RX positions so to ensure strong direct TX-RIS and RIS-RX LoS conditions, we assume free-space propagation for both the TX-RIS and RIS-RX links.
	
	\begin{rem}
		Since we only consider free-space propagation in this work, the outcomes could potentially apply also to sub-6 GHz links. However, we emphasize the mmWave case from a practical viewpoint since street-level implementation of transceiver nodes and RISs that can enable highly directional transmissions could be much more feasible in mmWave bands. This is attributed to the smaller packaging space needed in mmWave bands to achieve the same antenna gain compared with their sub-6 GHz counterparts. Consequently, we reckon our work as much more tailored to mmWave bands under practical deployment considerations.
		\end{rem}
	
	The RIS acts as a beamformer, which by adjusting the phase response of the RUs is capable of steering the beam at $\theta_{\text{r}}$, which is the RX direction. It consists of $N_x\times N_y$ RUs of size $d_x \times d_y$ and a controller that has perfect knowledge of the TX and RX positions. Each RU is an electrically-small low-gain element embedded on a substrate, with power radiation pattern that can be expressed as in~\cite{Tang_measurements_2021}
	\begin{align}
		\label{RIS_element_gain}
		G_{s}\left(\theta\right)= 4{\cos\left(\theta\right)}, \quad 0\le\theta<\pi/2.
	\end{align} 
Regarding the pattern of \eqref{RIS_element_gain}, it is reported that it is suitable for sub-wavelength RISs and it yields a good matching with respect to measurements conducted \cite{Wankai_Tang_Path_Loss_Modeling_Intelligent_Surfaces}, \cite{Tang_measurements_2021}.
	
	Due to the the fact that the TX-RIS and RIS-RX links are directional LoS links, they are deterministic. Moreover, it is assumed that the transmission power is $P_t$ and that the received signal is subject to additive white Gaussian noise with power~
	\begin{align}
		N_0=-174 + 10\log _{10} \left( W \right) + \mathcal{F}_{{\rm{dB}}},
	\end{align}
	where $\mathcal{F}_{{\rm{dB}}}$ is the noise figure in dB and $W$ is the transmission bandwidth~\cite{Ntontin_FD}.
	
	\section{RIS Illuminated area, SNR, and Optimal RIS placement}\label{S:SNR}
	
	In this section, we firstly derive the illuminated RIS area. Subsequently, we compute the received power assuming an RIS of arbitrary size. Moreover, we provide approximate expressions of the received power in the two cases of the RIS area being either much smaller or larger than the TX beam footprint. Finally, based on the corresponding approximate expressions, we analytically derive the optimal RIS placement for both cases. 
	
	\subsection{RIS's Illuminated Area}
	
	Since the main lobe of the TX antenna has a conical shape, its footprint in the RIS plane is an ellipse, according to the conic-section theory~\cite{hilbert1952geometry}.  
	\begin{lem}
		\label{transmit_energy_main_lobe}
		Under the pencil-beam transmission assumption, at least 97\% of the transmit energy is located within the first-null beamwidth (FNBW) of the TX beam, which we denote by $\phi_0$.
	\end{lem}
\begin{IEEEproof}
	The proof is provided in Appendix A. 
\end{IEEEproof}
According to Lemma~\ref{transmit_energy_main_lobe}, almost all of the transmit energy for pencil-beam transmissions is within the main lobe. Therefore, without loss of generality, we approximate the illuminated area at the RIS plane by the footprint corresponding to the particular lobe.
	
	\begin{lem}
		The two radii of the illuminated elliptic area at the RIS plane that corresponds to the FNBW can be obtained~as
		\begin{align}
			\alpha = \frac{\sin\left(\frac{\phi_0}{2}\right)}{\cos\left(\theta_i+\frac{\phi_0}{2}\right)}r_1
			\label{Eq:alpha}
		\end{align} 
		and 
		\begin{align}
			\beta = \alpha \sqrt{1-\epsilon^2},
			\label{Eq:beta}
		\end{align}
		where
		\begin{align}
			\epsilon = \frac{\sin\left(\theta_i\right)}{\cos\left(\frac{\phi_0}{2}\right)}.
			\label{Eq:epsilon}
		\end{align}
		Moreover, $r_1$  denotes the distance between the center of the TX and the center of the TX footprint at the RIS plane, while $\theta_i$ is the incident angle at the RIS center with respect to its broadside direction. 
	\end{lem} 
	\begin{IEEEproof}
		The proof is provided in Appendix B.  
	\end{IEEEproof}
	Based on~\eqref{Eq:alpha} and~\eqref{Eq:beta}, the TX main lobe footprint at the RIS plane can be evaluated~as
	\begin{align}
		S_i = \pi \alpha \beta.
		\label{Eq:Si}
	\end{align}
	
	The power that is reflected by the RIS is the one that falls~within 
	\begin{align}
		S = \min\left(S_i, S_s\right),
		\label{Eq:S}
	\end{align}  
	where $S_s$ denotes the RIS area. If $S_s \leq S_i$, only part of the power that falls within $S_i$ can be reflected towards the RX; thus, beam waste occurs. On the other hand, if $S_s > S_i$ only part of the RIS is used to reflect the incident electromagnetic wave\footnote{In the $S_s>S_i$ case, \eqref{Eq:S} corresponds to a very tight approximation due to the fact that under pencil-beam transmissions at least 97\% of the impinging power is included within $S_i$, according to Lemma~\ref{transmit_energy_main_lobe}.}. 
	
	Finally, we note that in the case $S_s>S_i$ there are some RUs in the perimeter of the ellipse that are partly illuminated, which would pose a challenge regarding how to adjust the amplitude and phase response of the particular RUs. However, by taking into account that the $S_s>S_i$ case would correspond to an illuminated RIS region of a relatively large size, ignoring those elements in the RU response adjustment process is not expected to have a notable effect on the resulting end-to-end performance.

	\subsection{End-to-end SNR}
	
	The following proposition returns a tight approximation for the received power. 
	\begin{prop}
		\label{Proposition_1}
		By adjusting the phase response of each of the RUs in a way that the received reflected signals are co-phased at the RX, which means that the received power, denoted by $P_R$, is maximized for certain $r_{1,h}$ and it can be evaluated~as\footnote{\eqref{received_power_RIS_Proposition_1} holds under the assumption of negligible mutual coupling among the RUs. Based on antenna theory, such an assumption approximately holds for adjacent RU distance equal to $\lambda/2$.}
		\begin{small}
			\begin{align}
				P_R &= \left(\frac{\lambda}{4\pi}\right)^4P_t\Gamma^2
				\left|\sum_{n=1}^{M}\sqrt{\frac{G_{t,n}G_{r,n}G_{s}{\left(\theta_{i,n}\right)}G_{s}{\left(\theta_{r,n}\right)}}{r_{1,n}^2 r_{2,n}^2}}\right|^2,
				\label{received_power_RIS_Proposition_1}
			\end{align}
		\end{small}
	\end{prop}where $\Gamma$ is the amplitude reflection coefficient that we consider is the same of all RUs,  $M$ is the number of illuminated RUs that are included within $S$. In the $S_s>S_i$ case, we consider that only the RUs that correspond to the FNBW are activated since that region contains at least 97\% of the impinging power, according to Lemma~\ref{transmit_energy_main_lobe}. Consequently, it holds that
\begin{equation}
	M=\frac{S}{d_xd_y}.
\end{equation}
$r_{1,n}$ and $r_{2,n}$ are the distances between the centers of the TX and RX antennas and the $n_{th}$ RU, respectively. In addition, $G_{t,n}$ and $G_{r,n}$ represent the TX and RX antenna gains corresponding to the same RU, respectively. Finally, $G_{s}{\left(\theta_{i,n}\right)}$ is the gain of the $n_{th}$ RU towards the TX antenna and $G_{s}{\left(\theta_{r,n}\right)}$ is its corresponding gain towards the RX antenna.
\begin{IEEEproof}
	The proof is provided in Appendix C.  
\end{IEEEproof}

In the special case in which $S_i>>S_s$, Proposition~\ref{Proposition_2} that follows presents a simplified closed-form expression for the received power.

\begin{prop}
	\label{Proposition_2}
	If $S_i>>S_s$, \eqref{received_power_RIS_Proposition_1} is reduced to
	\begin{small}
		\begin{align}
			P_R &=\left(\frac{\lambda}{4\pi}\right)^4
			\frac{P_t\Gamma^2\left(S_s\right)^2G_{t}^{max}G_{r}^{max}G_{s}{\left(\theta_{i}\right)}G_{s}{\left(\theta_{r}\right)}}{d_x^2d_y^2r_{1}^2 r_{2}^2},
			\label{received_power_RIS_Proposition_2}
		\end{align}
	\end{small}where $r_2$ denotes the distance between the center of the TX footprint at the RIS plane and the center of the RX. 
	\end{prop}
\begin{IEEEproof}
	We consider that $S_i>>S_s$ holds under far-field conditions, which means that the TX gain, RX gain, incident and departure RU gains together with the corresponding TX-RIS and RIS-RX distances are approximately independent of $n$.
	\end{IEEEproof}

\begin{lem}
	\label{Lemma_energy_step_function_HPBW}
	Under the pencil-beam transmission assumption, the amount of energy included within the FNBW can be tightly approximated to a level of at least 97\% by the amount of energy included within a step function with magnitude $G_{t}^{max}$ in the interval $\left[-\frac{\phi_{HPBW}}{2},\frac{\phi_{HPBW}}{2}\right]$. Furthermore, the particular amount of energy within the step function is at least equal to 94\% of the total impinging transmit energy at the RIS plane.
	\end{lem}
\begin{IEEEproof}
	The proof is provided in Appendix D.
	\end{IEEEproof}

In the special case in which $S_s\ge S_i$. Proposition~\ref{proposition_footprint_smaller_than_RIS} that follows presents a simplified closed-form expression for the received power.

\begin{prop}
	\label{proposition_footprint_smaller_than_RIS}
	If: i) $S_s\ge S_{i}$; ii) $G_{r,n}$ is aproximately independent of $n$ and equal to $G_{r}^{max}$; and iii) the incident and departure RU gains together with the corresponding TX-RIS and RIS-RX distances are approximately independent of $n$, $P_R$ is tightly approximated, by at least $94\%$ accuracy, as
	
	\begin{small}
		\begin{align}
			P_R &\approx \left(\frac{\lambda}{4\pi}\right)^4
			\frac{P_t\Gamma^2\left(S_{HPBW}\right)^2G_{t}^{max}G_{r}^{max}G_{s}{\left(\theta_{i}\right)}G_{s}{\left(\theta_{r}\right)}}{d_x^2d_y^2r_{1}^2 r_{2}^2},
			\label{received_power_RIS larger than the footprint}
		\end{align}
	\end{small}where $S_{HPBW}$ is the HPBW footprint of the main lobe on the RIS. $S_{HPBW}$ can be computed by the same process used in the computation of $S_i$, where in \eqref{Eq:alpha}, \eqref{Eq:beta}, and \eqref{Eq:epsilon}, $\phi_0$ should be replaced by $\phi_{HPBW}$.
	\end{prop}
\begin{IEEEproof}
	The proof of Proposition~\ref{proposition_footprint_smaller_than_RIS} is a direct result of Lemma~\ref{Lemma_energy_step_function_HPBW}.
	\end{IEEEproof}
In addition, by replacing $S_{HPBW}$ in \eqref{received_power_RIS larger than the footprint} with its corresponding expression, $P_R$ is further given by \eqref{Eq:P_R} at the top of the following page.

We note that the referred 94\% minimum approximation accuracy is achieved for the maximum HPBW of 15$^{\circ}$ needed for the transmission to be considered as pencil beam, based on Lemma~\ref{Lemma_energy_step_function_HPBW}. The smaller the HPBW is, the higher the accuracy becomes since more energy is included within the main-lobe region defined by the HPBW.

\begin{rem}
	Although the condition $S_i>>S_s$ can ensure that the RIS is located in the Fraunhofer  region of both the TX and RX antennas, which means that the impinging on the RIS electromagnetic wave can be considered as a plane wave, this does not necessarily hold in the $S_s\ge S_i$ case. In such a case, the requirements that the RX antenna gain is approximately constant over the illuminated RIS region and the incident and departure RU gains together with the corresponding TX-RIS and RIS-RX distances are approximately independent of $n$, under which \eqref{received_power_RIS larger than the footprint} holds, could be valid even if the phase of the impinging wave notably varies over the surface\footnote{The phase of an impinging wave on a surface should not vary by more than $\frac{\pi}{8}$ so that the wave is considered planar over the surface.}. Based on this, for the $S_s\ge S_i$ case the "far" condition in which the independency of the RX antenna gain, incident RU gain, departure RU gain, TX-RIS distance, and RIS-RX distance with respect to $n$ holds, is not necessarily equivalent to the Fraunhofer region, as it is also noted in \cite{Ellingson_Path_Loss_RIS} and \cite{RIS_Scaling_Laws_ERmil_Bjornson}.  
\end{rem}
Finally, the end-to-end SNR, which we denote by $\rho$, is given by dividing $P_R$ with $N_0$, i.e.
\begin{align}
	\label{end_to_end_SNR}
	\rho=\frac{P_R}{N_0}.
	\end{align}

	\begin{figure*}
		\begin{align}
			P_R&\approx\left(\frac{\lambda}{4\pi}\right)^4\frac{P_{t}\Gamma^2 G_{t}^{max} G_{r}^{max}G_{s}{\left(\theta_{i}\right)}G_{s}{\left(\theta_{r}\right)}} {d_x^2d_y^2}\left(\frac{r_{1}}{r_{2}}\right)^2
			\pi^2\frac{\sin^4\left(\frac{\phi_{HPBW}}{2}\right)}{\cos^4\left(\frac{\phi_{HPBW}}{2}+\theta_i\right)}
			\left(1-\frac{\sin^2\left(\theta_i\right)}{\cos^2\left(\frac{\phi_{HPBW}}{2}\right)}\right)
			\label{Eq:P_R}.
		\end{align}
		\hrulefill
	\end{figure*}

	
	\subsection{Optimal RIS Placement}
	
	 $r_1$, $r_2$, $\theta_i$, and $\theta_r$ can be expressed as 
	\begin{align}
		r_{1}&=\sqrt{r_{{1},{h}}^2+y_s^2+\left(h_s-h_{t}\right)^2},\\
		r_{2}&=\sqrt{\left(r_{h}-r_{1,{h}}\right)^2+y_s^2+\left(h_s-h_{r}\right)^2},\\
		\theta_i&=\tan^{-1}\left(\frac{\sqrt{r_{{1},{h}}^2+\left(h_s-h_{t}\right)^2}}{y_s}\right)
	\end{align}
	and
	\begin{align}
		\theta_r&=\tan^{-1}\left(\frac{\sqrt{\left(r_{{1},{h}}-r_h\right)^2+\left(h_s-h_{r}\right)^2}}{y_s}\right),
	\end{align}for $y_s>0$. Next, for the $S_i>>S_s$ and $S_s \ge S_i$ cases we determine the $r_{1,h}$ that maximizes the end-to-end SNR. 
	
	\subsubsection{$S_{i}>>S_s$ case}
	\begin{prop}
		\label{Proposition_Optimal_RIS_placement_small}
		The optimum TX-RIS horizontal distance, denoted by $r_{1,h}^{*}$, that maximizes the end-to-end SNR can be obtained by taking the 1st derivative of $\rho$ with respect to $r_{1,h}$ and setting it equal to 0. This yields 
			\begin{align}
				\label{3rd_degree_polynomial_RIS_small}
			a^{\left(1\right)} r_{1,h}^3 + b^{\left(1\right)} r_{1,h}^2 + c^{\left(1\right)} r_{1,h} + d^{\left(1\right)} = 0,
		\end{align} 
	\end{prop}where
\begin{align}
	&a^{\left(1\right)}=6,\\
	&b^{\left(1\right)}=-9r_{h},\\
	&c^{\left(1\right)}=3\left(2y_s^2+r_{h}^2+(h_s-h_t)^2+(h_s-h_r)^2\right),\\
    &d^{\left(1\right)}=-3r_{h}\left(y_s^2+\left(h_s-h_t\right)^2\right).
\end{align}
	\begin{IEEEproof}
		The proof is provided in Appendix E.
	\end{IEEEproof}


\subsubsection{$S_s\ge S_i$ case}

\begin{prop}
	\label{Proposition_Optimal_RIS_placement_large}
	 By by taking the 1st derivative of $\rho$ with respect to $r_{1,h}$ and setting it equal to 0, $r_{1,h}^{*}$ can approximated by \eqref{solution_RIS_large}, given at the top of the next page. 
	\begin{figure*}          
	\begin{align}
		\label{solution_RIS_large}
		r_{1,h}^{*}\approx\frac{-\left(h_s-h_t\right)^2+r_h^2+\left(h_s-h_r\right)^2+\sqrt{\left(\left(h_s-h_t\right)^2-r_h^2-\left(h_s-h_r\right)^2\right)^2+4r_h^2\left(y_s^2+\left(h_s-h_t\right)^2\right)}}{2r_h}.
	\end{align}
\hrulefill
\end{figure*}
\end{prop}
	\begin{IEEEproof}
	The proof is provided in Appendix F.
\end{IEEEproof}
The tightness of \eqref{solution_RIS_large} with respect to the exact value of $r_{1,h}^{*}$ is validated in Section~\ref{S:Res} by means of simulations.

	\section{Numerical Results \& Design Guidelines}\label{S:Res}
	The aim of this section is twofold: i) to validate, by means of simulations, Proposition~\ref{Proposition_2}, Proposition~\ref{proposition_footprint_smaller_than_RIS}, and the analytical frameworks for the computation of $r_{1,h}^{*}$ based on \eqref{3rd_degree_polynomial_RIS_small} and \eqref{solution_RIS_large};) and ii) to provide design guidelines based on the resulting trends for various configurations. 
	
    \subsection{Results}
	
	We consider the parameters of Table~\ref{Parameter_values}. 
	
	\begin{table}[h]
		\label{Parameter_values}
		\caption{Parameter values used in the simulation.} 
		\centering 
		\scalebox{0.8}{
			\begin{tabular}{| c | c | c | } 
				\hline
				Parameter & Value \\[0.5ex]
				\hline
				\hline
				$f$& $140$ GHz \\ [0.5ex]
				\hline
				$P_{t}$& $1$ W\\ [0.5ex]
				\hline
				$W$& $2$ GHz \\ [0.5ex]
				\hline
				$\mathcal{F}_{{\rm{dB}}}$& $10$ dB\\[0.5ex]
				\hline
				$d_x$, $d_y$& $\lambda/2$ \\[0.5ex]
				\hline
				$h_{s}$& $12$ m \\ [0.5ex]
				\hline
				$D_{t}$& $15$ cm \\ [0.5ex]
				\hline
				$e_{t}$, $e_{r}$ & $0.7$ \\ [0.5ex]
				\hline
				$\Gamma$ & $0.9$ \\ [0.5ex]
				\hline
		\end{tabular}}
		\label{Parameter_values} 
	\end{table}
	
	\subsubsection{Validation of Proposition~\ref{Proposition_2}}
	
	\begin{figure}
		\label{RIS_size_small_effect_y_s}
		\centering
		{\includegraphics[width=3.6in,height=2.4in]{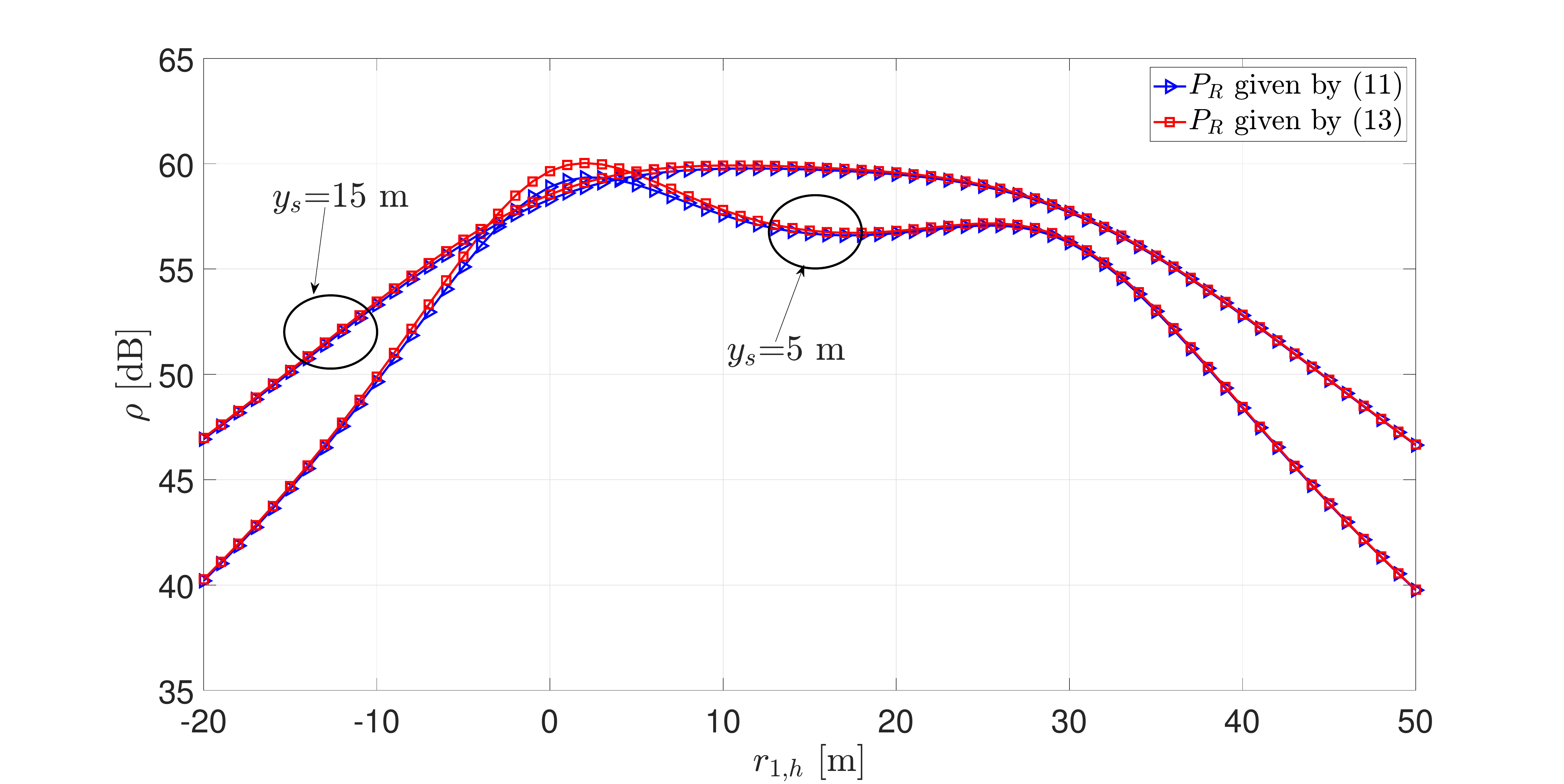}}\\
		(a) $h_t=6$ m, $h_r=3$ m.\\
		{\includegraphics[width=3.6in,height=2.4in]{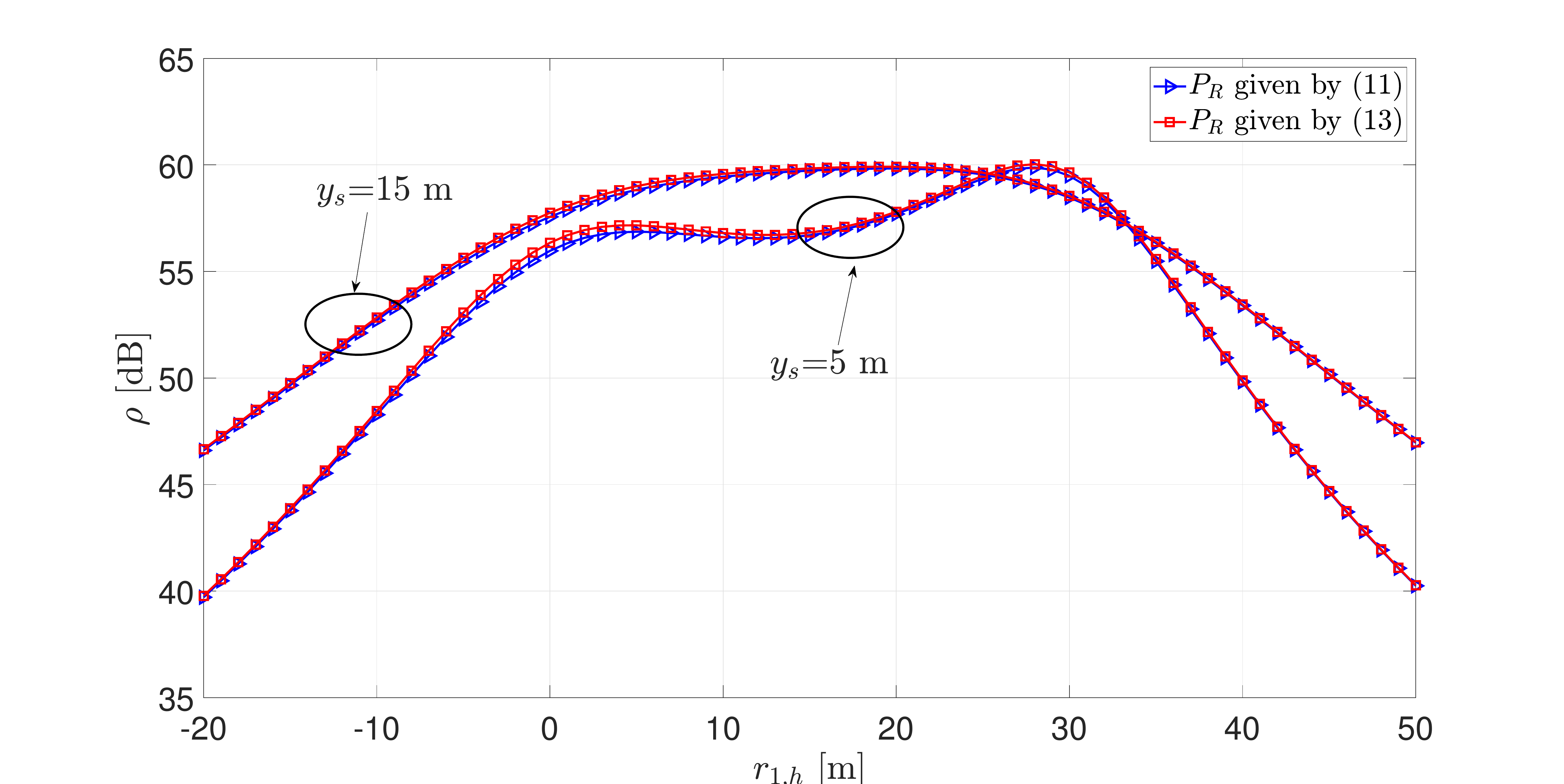}}\\
		(b)  $h_t=3$ m, $h_r=6$ m. \\
		{\includegraphics[width=3.6in,height=2.4in]{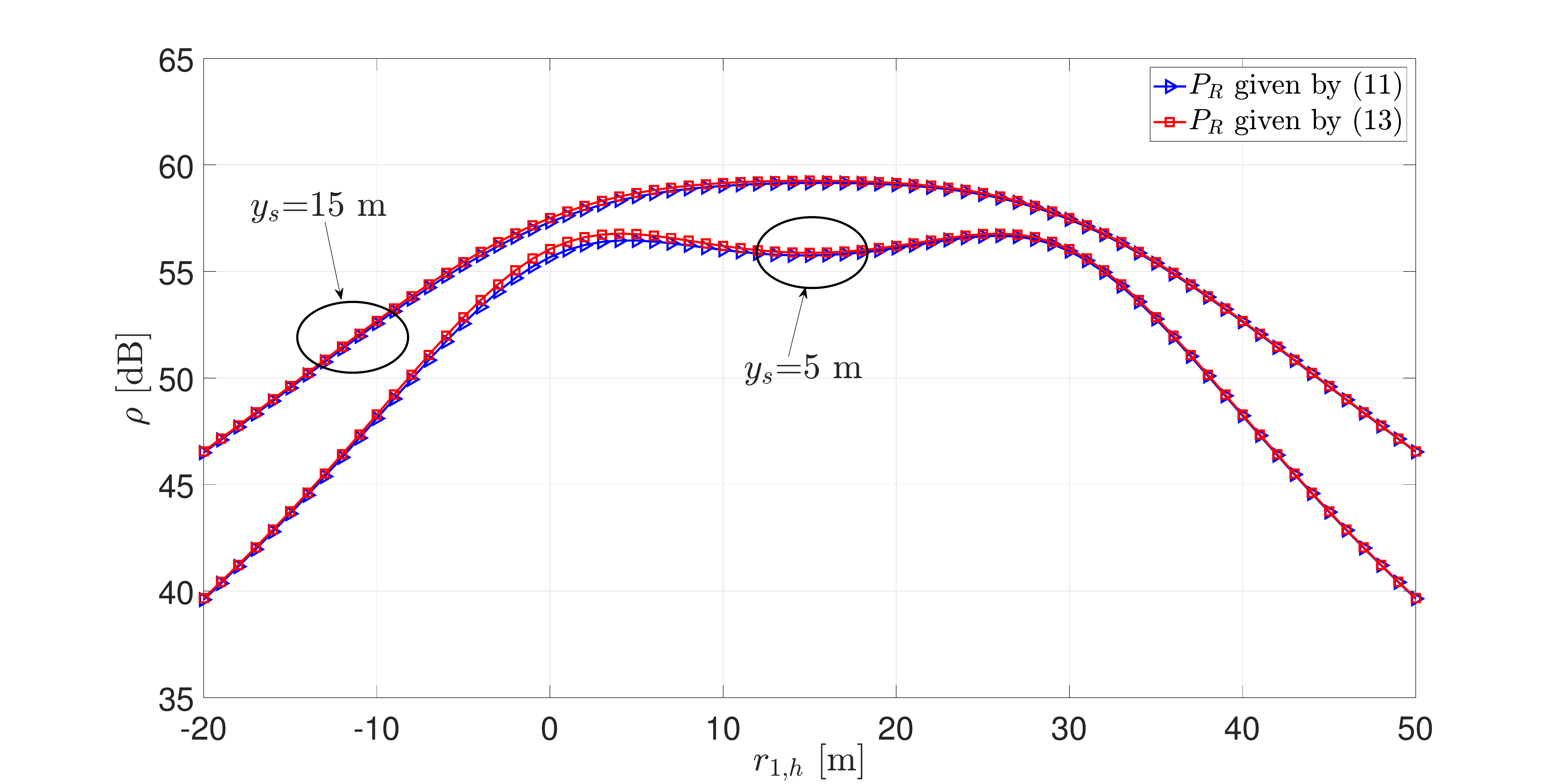}}\\
		(c)  $h_t=3$ m, $h_r=3$ m. \\
		\caption{$\rho$ vs. $r_{1,h}$ for $D_r=3$ cm, $r_h=30$ m, and $S_s=0.012$ $m^2$.}
		\label{RIS_size_small_effect_y_s}
	\end{figure}

	\begin{figure}
		\label{RIS_size_large_effect_y_s}
		\centering
		{\includegraphics[width=3.6in,height=2.4in]{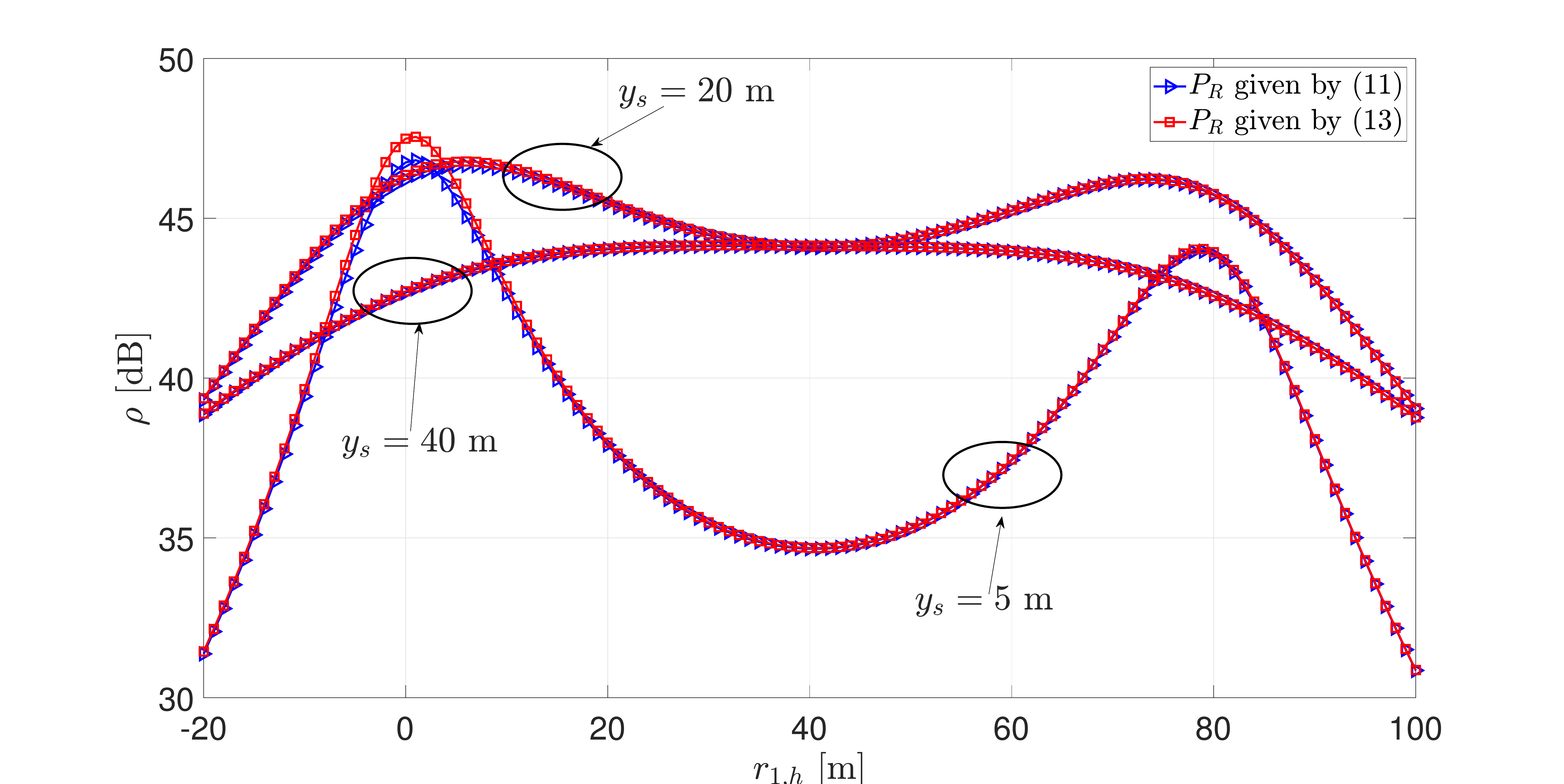}}\\
		(a) $h_t=6$ m, $h_r=3$ m.\\
		{\includegraphics[width=3.6in,height=2.4in]{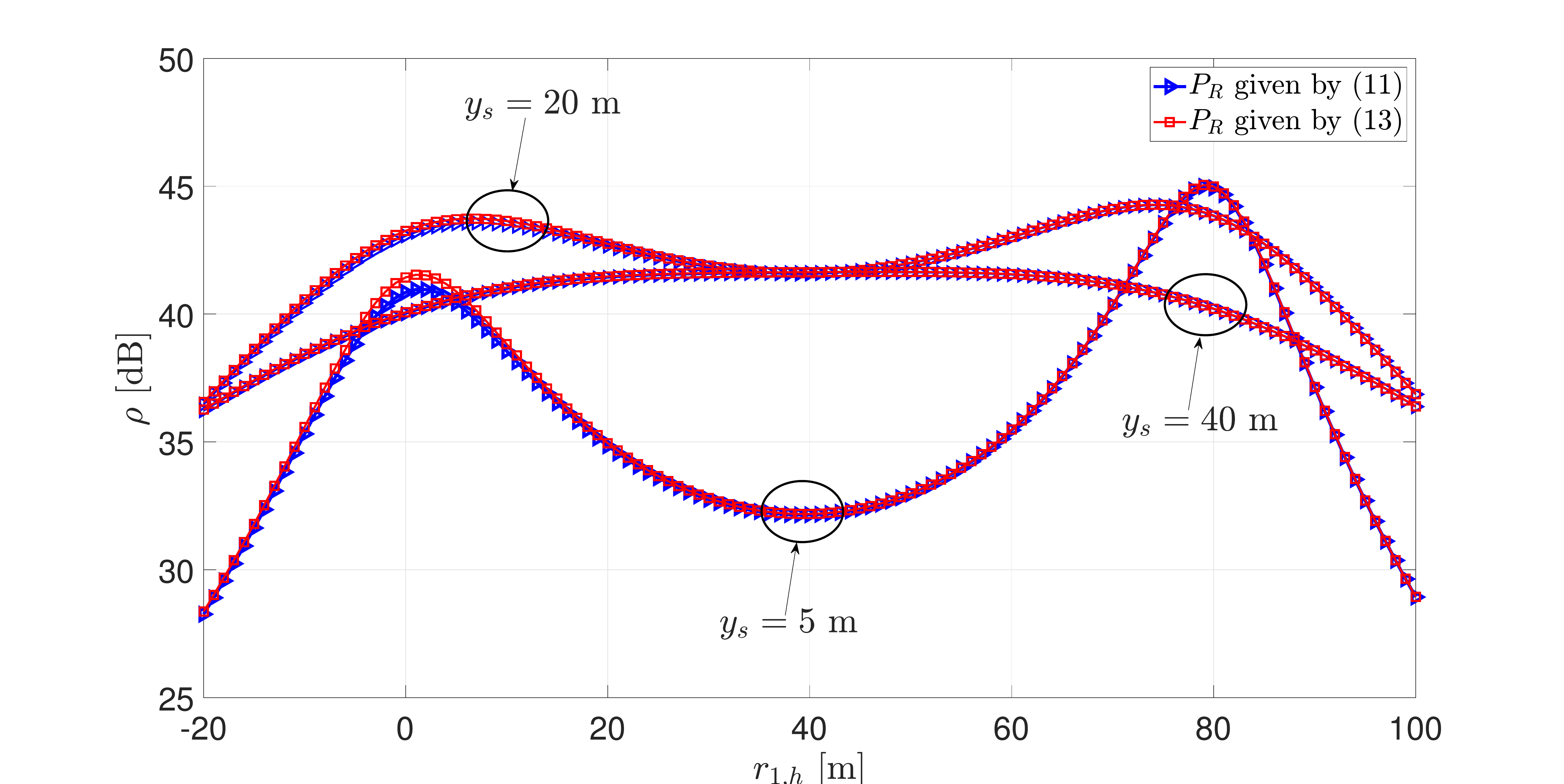}}
	    (b)  $h_t=3$ m, $h_r=6$ m. \\
		\caption{$\rho$ vs. $r_{1,h}$ for $D_r=3$ cm, $r_h=80$ m, and $S_s=0.012$ $m^2$.}
		\label{RIS_size_large_effect_y_s}	
	\end{figure}

\begin{figure}
	\centering
	{\includegraphics[width=3.6in,height=2.4in]{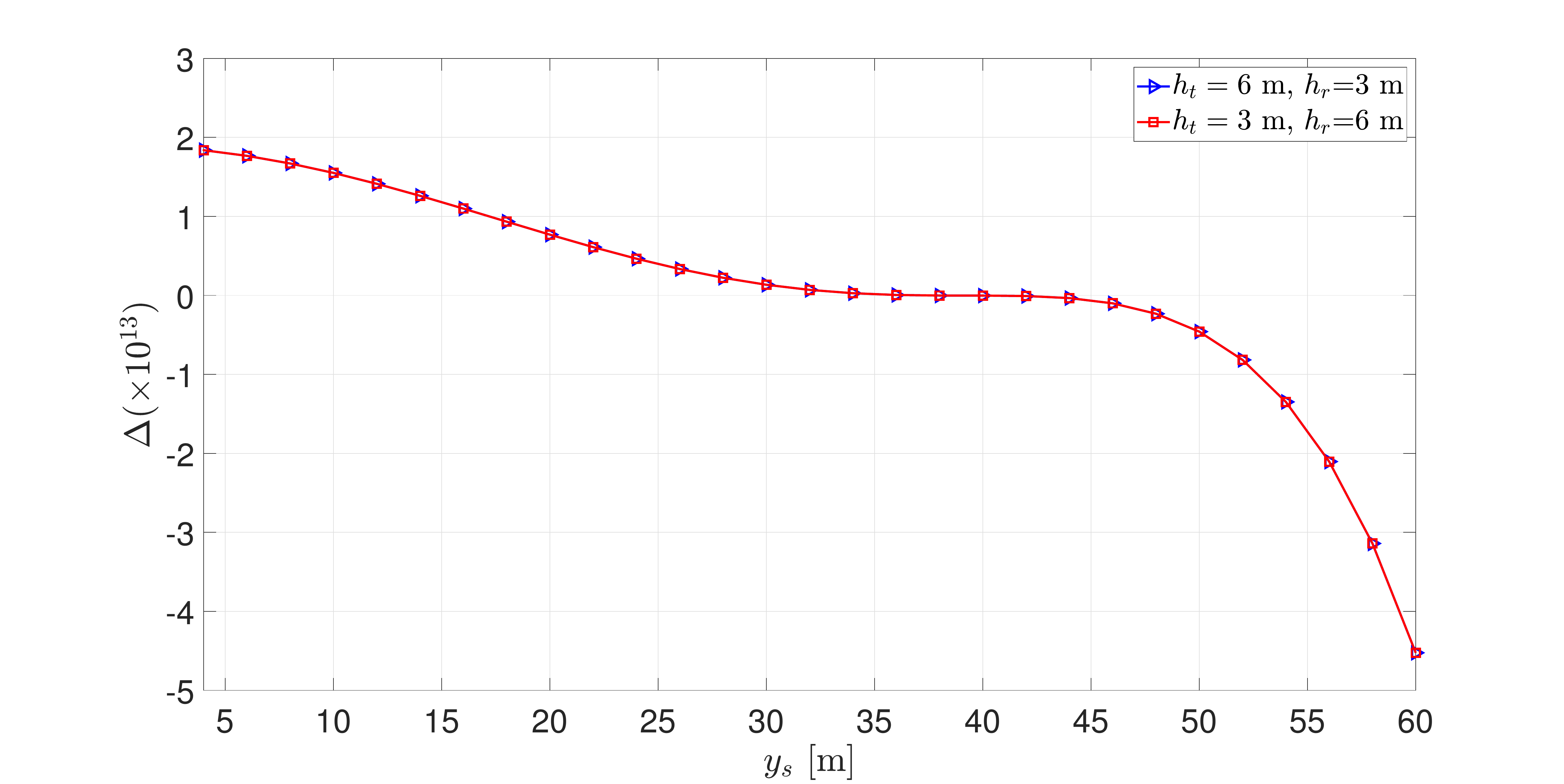}}
	\caption{Discriminant of \eqref{3rd_degree_polynomial_RIS_small} for $h_t=6$ m, $h_r=3$ m, $D_r=3$ cm, and $r_h=80$ m.}
	\label{Fig:Discriminant}	
\end{figure}
	
	\begin{figure}
		\centering
		{\includegraphics[width=3.6in,height=2.4in]{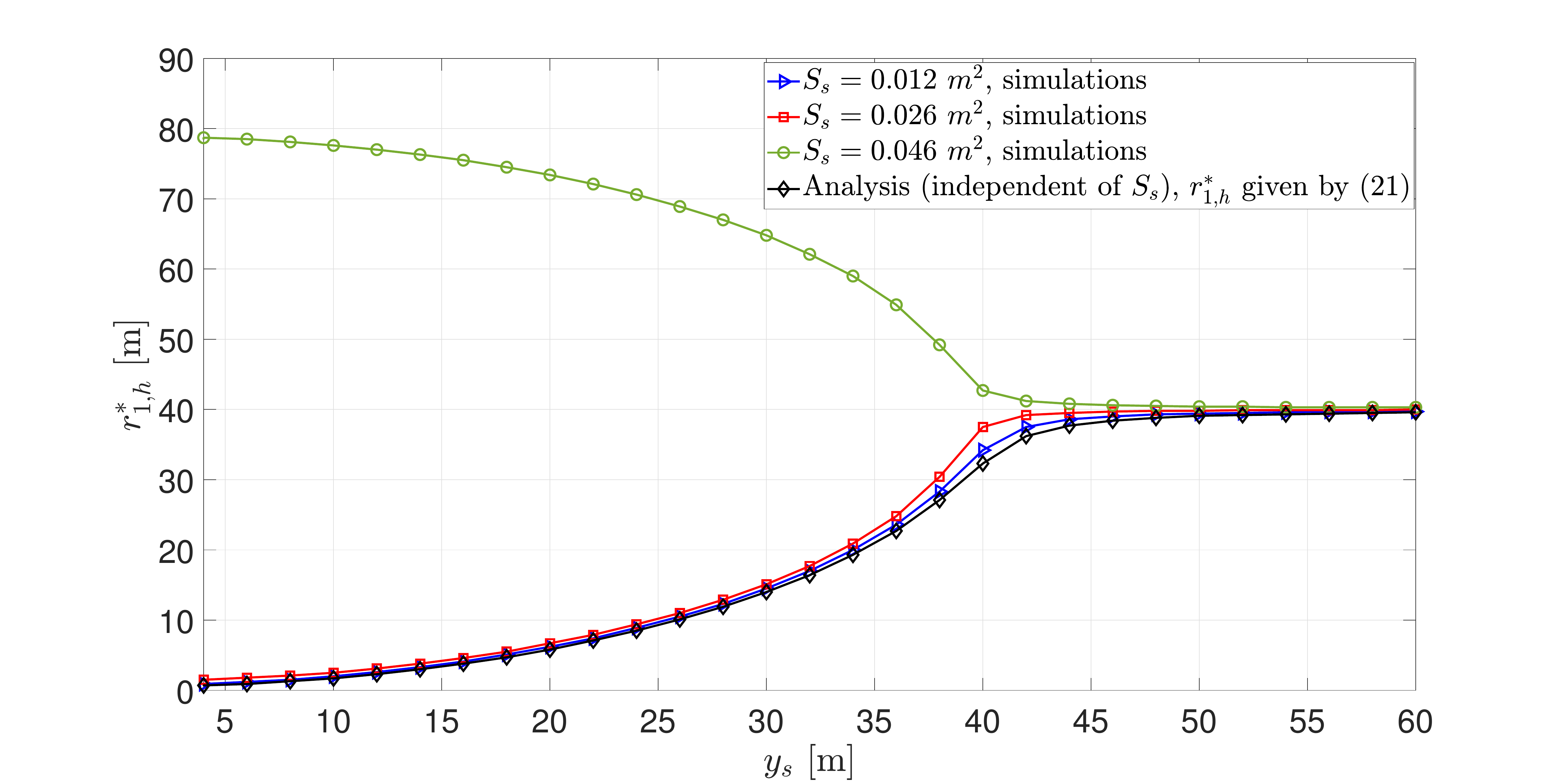}}
		\caption{$r_{1,h}^{*}$ vs. $y_s$ for $h_t=6$ m, $h_r=3$ m, $D_r=3$ cm, and $r_h=80$ m.}
		\label{Fig:r_1_h_optimal_vs_y_s}	
	\end{figure}
	
	\begin{figure}
		\centering
		{\includegraphics[width=3.6in,height=2.4in]{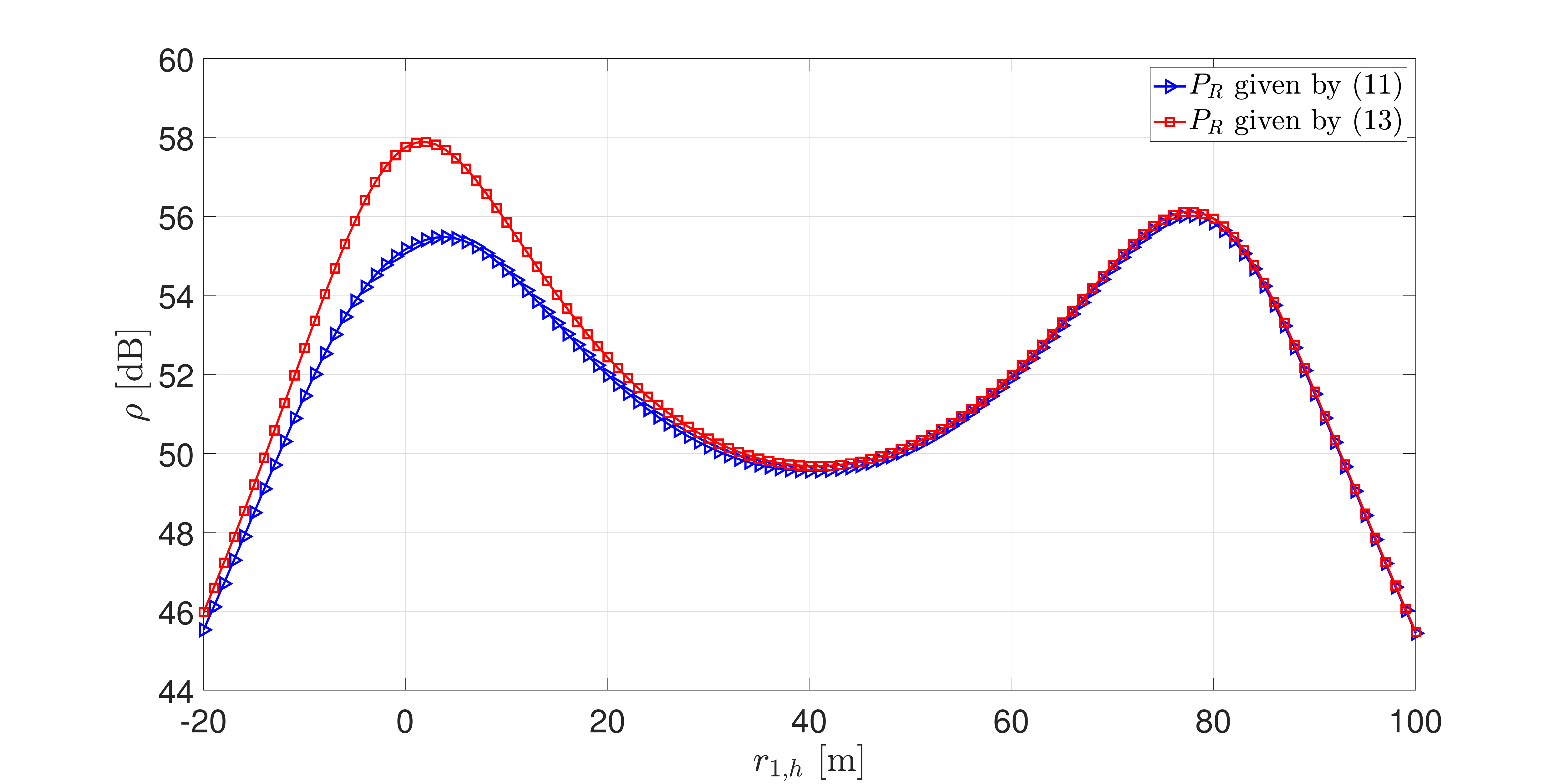}}
		\caption{$\rho$ vs. $r_{1,h}$ for $h_t=6$ m, $h_r=3$ m, $D_r=3$ cm, $r_h=80$ m, $y_s=10$ m, and $S_s=0.046$ $m^2$.}
		\label{Fig:rho_vs_r_1_h_Size_RIS_0046}	
	\end{figure}
	
	As far as the validation of Proposition~\ref{Proposition_2} is concerned, in Fig.~\ref{RIS_size_small_effect_y_s} we depict the exact, based on \eqref{received_power_RIS_Proposition_1}, and closed-form, based on \eqref{received_power_RIS_Proposition_2},  $\rho$ vs. $r_{1,h}$ curves for $S_s=0.012$ $m^2$, $D_r=3$ cm, $r_h=30$ m, and different values of $h_t$, $h_r$, and $y_s$. We note that for the considered value of $D_t$ it holds that $\phi_{HPBW}=1.25^{\circ}$, which means that a highly pencil beam transmission is enacted. In addition, for the examined scanning range of $r_{1,h}$ the minimum value of $S_i$ is equal to $0.09$ $m^2$ and $0.69$ $m^2$ for $y_s=5$ m and $y_s=15$ m, respectively. Hence, it holds that $S_i>>S_s$ throughout the considered $r_{1,h}$ range in both cases. 
	
	As we observe from Fig.~\ref{RIS_size_small_effect_y_s}, there is a relatively good match between the exact and closed-form expressions of $\rho$, which validates \eqref{received_power_RIS_Proposition_2}. In addition, we observe that for $y_s=5$ m $\rho$ is maximized when the RIS is closer to the TX, closer to the RX, and either closer to the TX or the RX for $h_t>h_r$, $h_t<h_r$, and $h_t=h_r$, respectively. In addition, $\rho$ is minimized if the RIS is placed close to the middle of the TX-RX distance. However, there is a relatively small variation among the two local optima of $\rho$ and the local minimum near the middle of the TX-RX distance. On the other hand, for $y_s=15$ m we observe that the optimal placement of the RIS has moved closer to the middle of the TX-RX distance in the $h_t>h_r$ and $h_t<h_r$ cases. This indicates that the higher $y_s$ is, the less pronounced the effect of the height difference between the TX and RX antennas is on the optimal RIS placement. Consequently, $\rho$ is maximized when the RIS is placed near to the middle of the TX-RX distance, as we observe from Fig.~\ref{RIS_size_small_effect_y_s}.

	To further validate the observed trends mentioned in the previous paragraph, in Fig.~\ref{RIS_size_large_effect_y_s} we depict the $\rho$ vs. $r_{1,h}$ curves for a much larger $r_h$ distance and 3 values of $y_s$. As we observe from Fig.~\ref{RIS_size_large_effect_y_s}, for small values of $y_s$ the variation among the two local optima of $\rho$ and its local minimum is notable. In particular, a 12.5 dB difference is observed between the value that maximizes $\rho$ and the one that minimizes it near the middle of the TX-RX distance. Furthermore, Fig.~\ref{RIS_size_large_effect_y_s} again reveals that the two local optima of $\rho$ do not differ substantially in value. As $y_s$ increases, the corresponding difference diminishes. More specifically, as $y_s$ increases \eqref{3rd_degree_polynomial_RIS_small} moves from having 3 real roots, which results in 2 local optima and 1 local minimum, to having only 1 real root for sufficiently large $y_s$. 
	
	The trends of Fig.~\ref{RIS_size_large_effect_y_s} can be validated by observing how the discriminant of the 3rd degree polynomial of \eqref{3rd_degree_polynomial_RIS_small}, which we denote by $\Delta$, varies with respect to $y_s$. Regarding this, in  Fig.~\ref{Fig:Discriminant} we depict $\Delta$ with respect to $y_s$. As we observe from Fig.~\ref{Fig:Discriminant}, as $y_s$ increases $\Delta$ moves from positive values, which means that \eqref{3rd_degree_polynomial_RIS_small} has 3 real roots, to negative values, which means that only one real root exists. 
	
	Finally, in order to examine the effect of $S_s$ on the optimal RIS placement, in Fig.~\ref{Fig:r_1_h_optimal_vs_y_s} we compare the exact $r_{1,h}^{*}$ vs. $y_s$ curves, based on $\eqref{received_power_RIS_Proposition_1}$, with the closed-form analytically obtained one, based on $\eqref{3rd_degree_polynomial_RIS_small}$, for $r_h=80$ m and 3 values of $S_s$. Although there is a relatively close match of the obtained by simulations exact $r_{1,h}^{*}$ values with the analytical one for $S_s=0.012$ $m^2$ and $S_s=0.026$ $m^2$, there is a substantial discrepancy between them in the $S_s=0.046$ $m^2$ case for low-to-moderate $y_s$ values. This is justified by the fact that for the particular $y_s$ values it does not hold that the minimum value of $S_i$ in in the considered scanning range of $r_{1,h}$ is much larger than $S_s$. For instance, for $y_s=10$ m the minimum value of $S_i$, which occurs for $r_{1,h}=0$, is equal to $0.15$ $m^2$. Hence, it does not hold that $S_i>>S_s$ throughout the $r_{1,h}$ scanning range, which results in \eqref{received_power_RIS_Proposition_2} providing inaccurate results. To substantiate the latter, in Fig.~\ref{Fig:rho_vs_r_1_h_Size_RIS_0046} we illustrate the $\rho$ vs. $r_{1,h}$ curve for $S_s=0.046$ $m^2$ and $y_s=10$ m. As we observe from Fig.~\ref{Fig:rho_vs_r_1_h_Size_RIS_0046}, there is a substantially discrepancy between the exact and the closed-form $\rho$ curves for $r_{1,h}=0$. As aforementioned, in the particular position $S_i$ exhibits its smallest value, equal to 0.15 $m^2$, and, hence, it holds that $S_s/S_i=0.31$, which is smaller than 1, but not notably smaller. The latter is the requirement for \eqref{received_power_RIS_Proposition_2} to hold.

	\subsubsection{Validation of Proposition~\ref{proposition_footprint_smaller_than_RIS}}   
	
    	\begin{figure}
    	\centering
    	{\includegraphics[width=3.6in,height=2.4in]{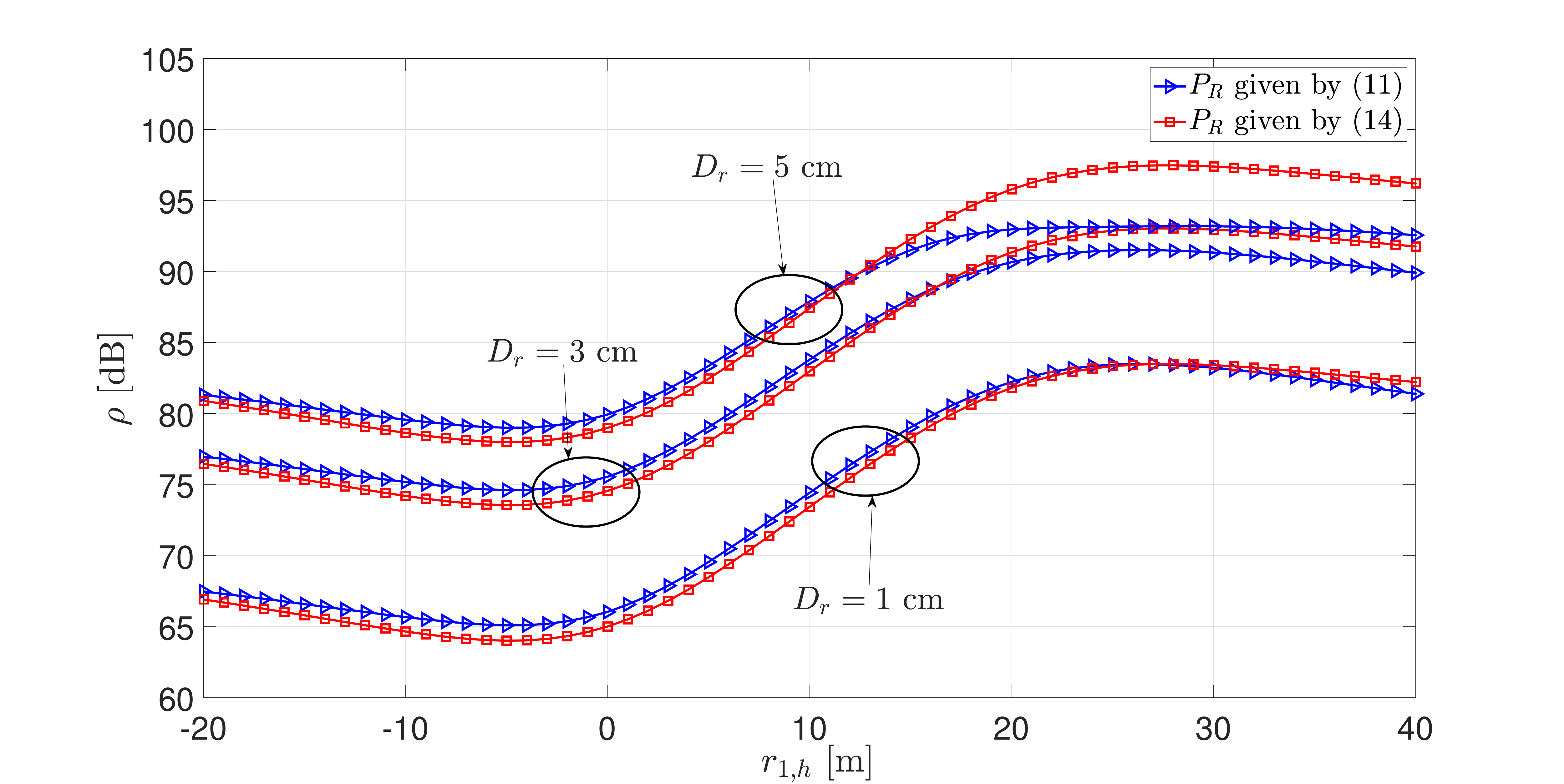}}\\
    	(a) $h_t=6$ m, $h_r=3$ m.\\
    	{\includegraphics[width=3.6in,height=2.4in]{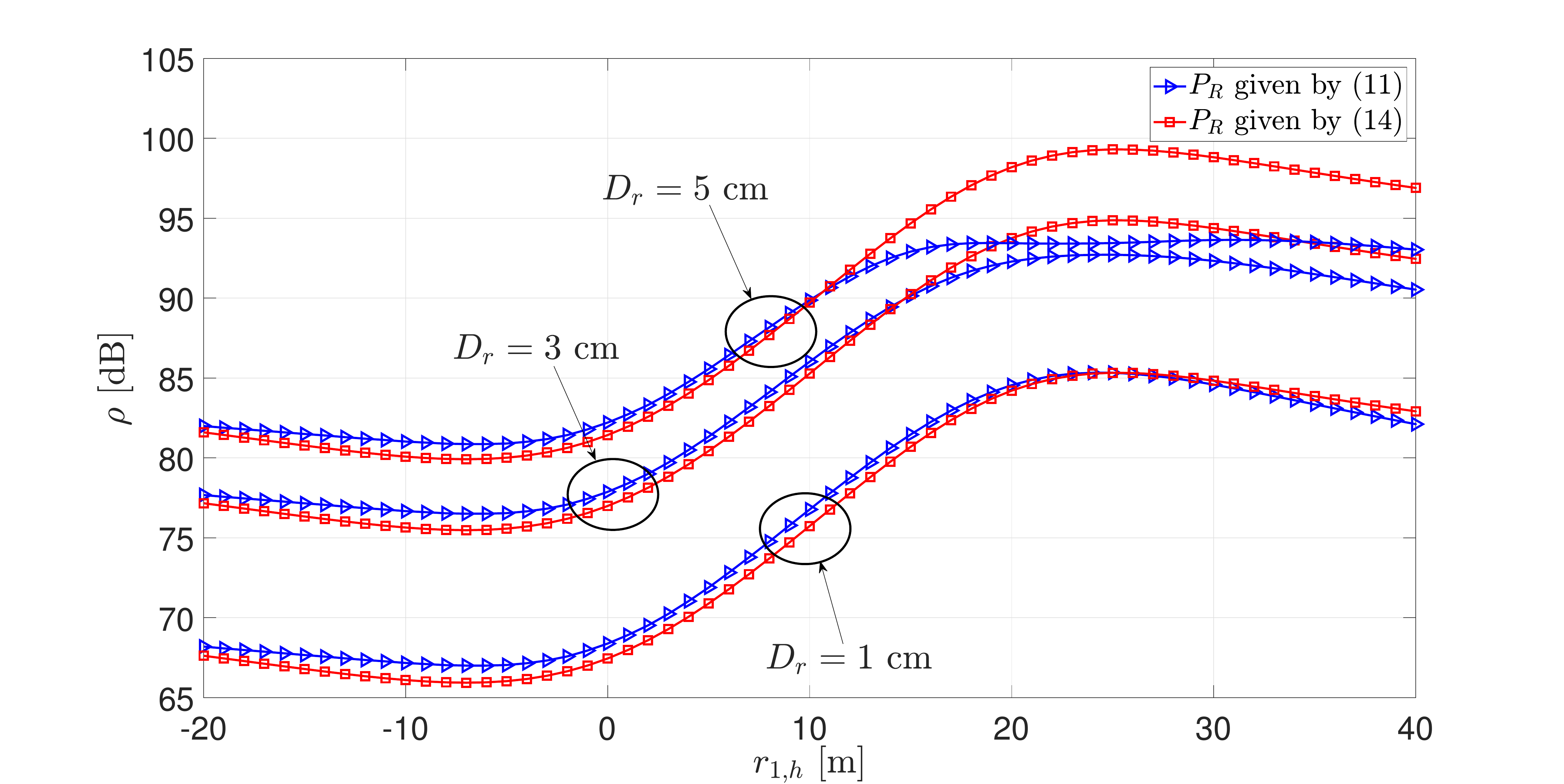}}
    	(b)  $h_t=3$ m, $h_r=6$ m. \\
    	\caption{$\rho$ vs. $r_{1,h}$ for $r_h=20$ m, $y_s=10$ m, and $S_s>S_i$.}
    	\label{Fig:rho_vs_r_1_h_RIS_large}	
    \end{figure}

	\begin{figure}
	\centering
	{\includegraphics[width=3.6in,height=2.4in]{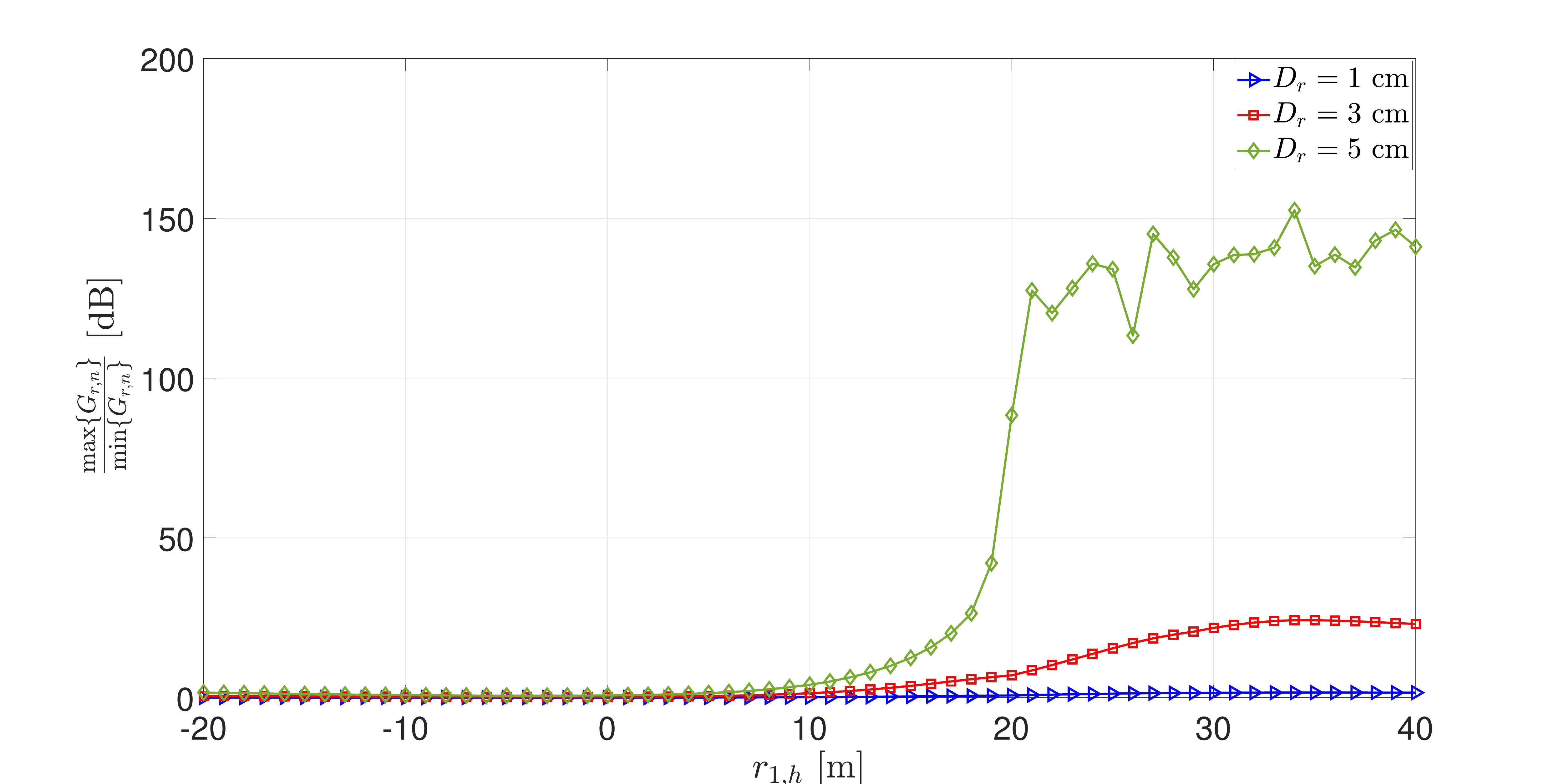}}
	\caption{$\frac{\max\left\{G_{r,n}\right\}}{\min\left\{G_{r,n}\right\}}$ vs. $r_{1,h}$ for $h_t=6$ m, $h_r=3$ m, $r_h=20$ m, $y_s=10$ m, and $S_s>S_i$.}
	\label{Fig:Max_min_gain_difference_RX_antenna}	
\end{figure}

	\begin{figure}
	\centering
	{\includegraphics[width=3.6in,height=2.4in]{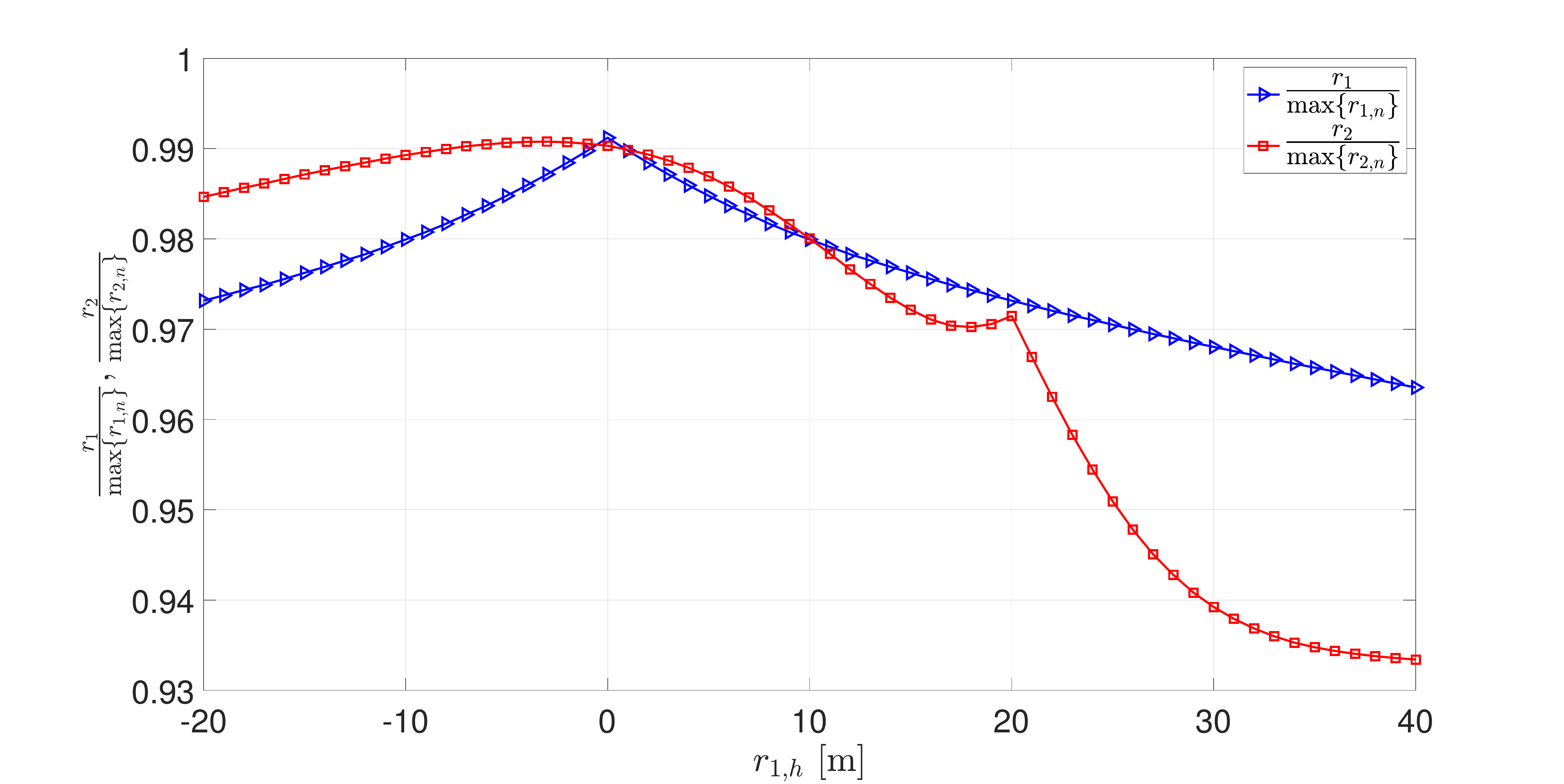}}
	\caption{$\frac{r_1}{\max\left\{r_{1,n}\right\}}$ and $\frac{r_2}{\max\left\{r_{2,n}\right\}}$ vs. $r_{1,h}$ for $h_t=6$ m, $h_r=3$ m, $D_r=5$ cm, $r_h=20$ m, $y_s=10$ m, and $S_s>S_i$.}
	\label{Fig:variation_of_distance_over_the_RIS}	
\end{figure}

	\begin{figure}
	\centering
	{\includegraphics[width=3.6in,height=2.4in]{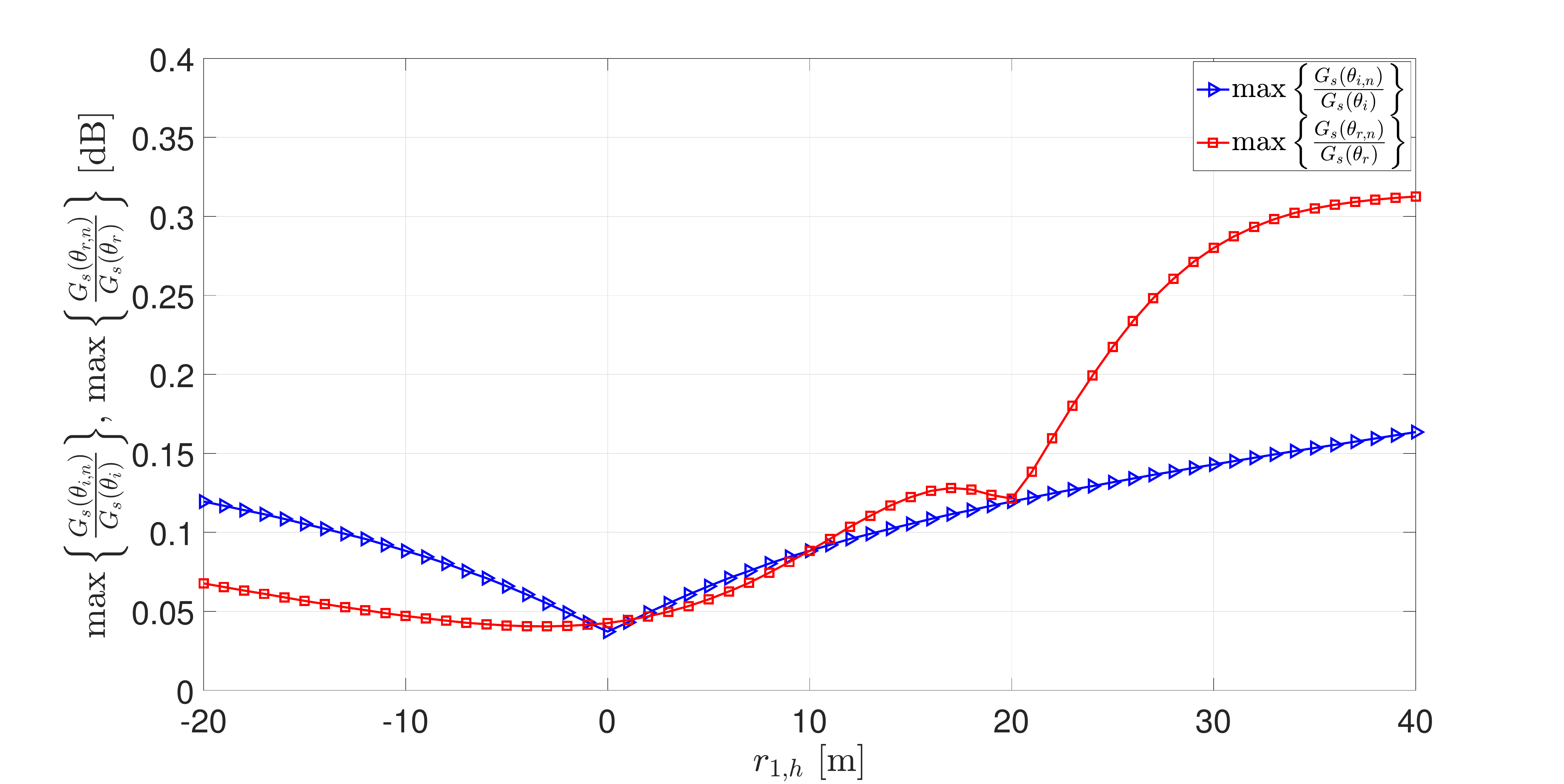}}
	\caption{$\max\left\{\frac{G_s\left(\theta_{i,n}\right)}{G_s\left(\theta_i\right)}\right\}$ and $\max\left\{\frac{G_s\left(\theta_{r,n}\right)}{G_s\left(\theta_r\right)}\right\}$ vs. $r_{1,h}$ for $h_t=6$ m, $h_r=3$ m, $D_r=5$ cm $r_h=20$ m, $y_s=10$ m, and $S_s>S_i$.}
	\label{Fig:variation_of_gains_RUs_over_the_RIS}	
\end{figure}

\begin{figure}
	\centering
	{\includegraphics[width=3.6in,height=2.4in]{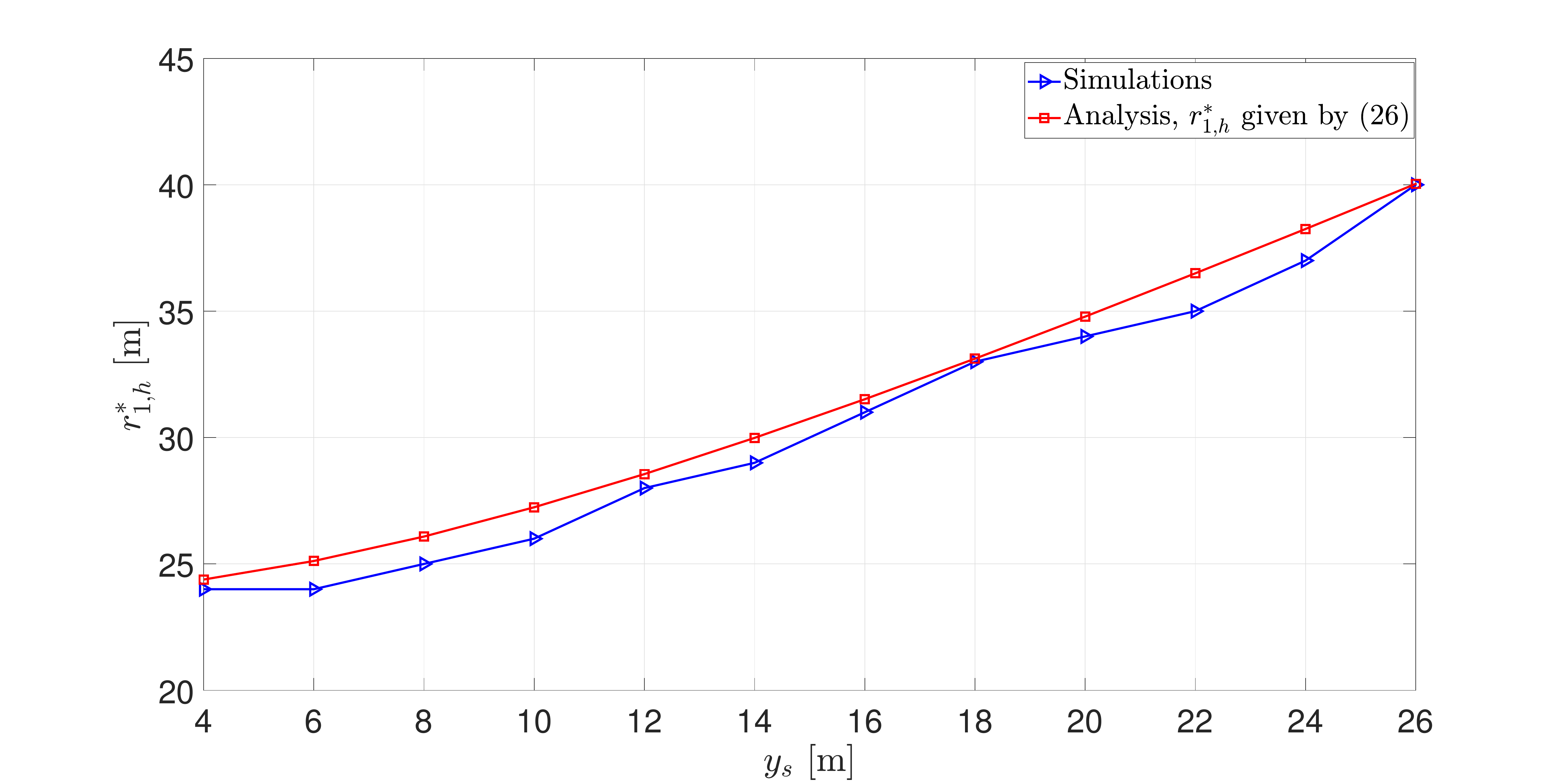}}
	\caption{$r_{1,h}^{*}$ vs. $y_s$ for $h_t=6$ m, $h_r=3$ m, $r_h=20$ m, $D_r=1$ cm, and $S_s>S_i$.}
	\label{Fig:optimal_point_vs_ys_RIS_large}	
\end{figure}

As far as the validation of Proposition~\ref{proposition_footprint_smaller_than_RIS} is concerned, in Fig.~\ref{Fig:rho_vs_r_1_h_RIS_large} we depict the $\rho$ vs. $r_{1,h}$ curves for $r_h=20$ m, $y_s=10$ m, and different $D_r$ under the assumption that $S_s>S_i$ throughout the examined $r_{1,h}$ range. In particular, $S_i$ takes its largest value for $r_{1,h}=40$ m, which is equal to 7.97 $m^2$. Outdoor objects where such large RIS surfaces can be mounted are, for instance, the facades of buildings. As we see from Fig.~\ref{Fig:rho_vs_r_1_h_RIS_large}, the RIS should be placed closer to the RX so that $\rho$ is maximized. 

In addition, from  Fig.~\ref{Fig:rho_vs_r_1_h_RIS_large} we further observe that the higher $D_r$ is, the larger the deviation of the exact value of $\rho$, based on $\eqref{received_power_RIS_Proposition_1}$, with its closed-form counterpart, based on \eqref{received_power_RIS larger than the footprint}, is for larger $r_{1,h}$ values. This discrepancy is attributed to the fact that as $D_r$ increases the footprint of the main lobe of the RX antenna on the RIS reduces and, consequently, $G_{r,n}$ cannot be considered approximately constant for the illuminated RIS region, as it is required so that \eqref{received_power_RIS larger than the footprint} holds with high accuracy. This is more pronounced for larger $r_{1,h}$ values due to the larger illuminated region on the RIS and its closer distance with respect to the RX antenna. 

To substantiate our claim about the larger variation of $G_{r,n}$ across the illuminated RIS region as $D_r$ increases, in Fig.~\ref{Fig:Max_min_gain_difference_RX_antenna} we illustrate how the ratio $\frac{\max\left\{G_{r,n}\right\}}{\min\left\{G_{r,n}\right\}}$ varies with respect to $r_{1,h}$ for 3 values of $D_r$. As we observe, for $D_r=1$ cm $G_{r,n}$ remains almost constant across the illuminated RIS region, but it starts varying as $D_r$ increases. In particular, it varies substantially for $D_r=5$ cm in the $r_{1,h}>10$ m region, which justifies the increasing discrepancy in the particular region between the exact and closed-form results as $r_{1,h}$ increases. 

On the other hand, from Fig.~\ref{Fig:variation_of_distance_over_the_RIS}, which depicts the ratios $\frac{r_1}{\max\left\{r_{1,n}\right\}}$ and $\frac{r_2}{\max\left\{r_{2,n}\right\}}$ with respect to $r_{1,h}$, we see that throughout the $r_{1,h}$ scanning range there are small variations of the distances between the TX (RX) antenna center and the $n_{th}$ RU, which indicates that $r_{1,n}$ and $r_{2,n}$ can be closely approximated by $r_1$ and $r_2$, respectively, for each $n$. Furthermore, small variations are also observed for $G_s\left(\theta_{i,n}\right)$ and $G_s\left(\theta_{r,n}\right)$ with respect to $G_s\left(\theta_{i}\right)$ and $G_s\left(\theta_{r}\right)$, respectively, across the $r_{1,h}$ scanning range, as Fig.~\ref{Fig:variation_of_gains_RUs_over_the_RIS} depicts. Consequently, from Fig.~\ref{Fig:Max_min_gain_difference_RX_antenna}, Fig.~\ref{Fig:variation_of_distance_over_the_RIS}, and Fig.~\ref{Fig:variation_of_gains_RUs_over_the_RIS}, we conclude that the variations of $G_{r,n}$ over the illuminated RIS region are the reason for the large discrepancy between the exact numerical and theoretical results in the $r_{1,h}>10$ m region for $D_r=5$ cm that is observed in Fig.~\ref{Fig:rho_vs_r_1_h_RIS_large}.

Finally, to validate the close match of $r_{1,h}^{*}$ obtained by \eqref{solution_RIS_large} with the exact value obtained by simulations, in Fig.~\ref{Fig:optimal_point_vs_ys_RIS_large} we depict the variation of both the analytically obtained $r_{1,h}^{*}$ and the one by simulations with respect to $y_s$ for $D_r=1$ cm. As we see, there is a relatively close match of the two curves, which validates \eqref{solution_RIS_large}.  In addition, the higher $y_s$ is the larger $r_{1,h}^{*}$ becomes.

\subsection{Design Guidelines}

Let us now recapitulate on the main outcomes regarding the $S_i>>S_s$ and $S_i \le S_s$ cases, according to the presented numerical results, that can be used by the system designer so that the optimal RIS placement is achieved based on the system parameters.

\subsubsection{$S_i>>S_s$ case}

\begin{itemize}
	\item For relatively small values of $y_s$ compared with $r_h$, there are two positions, one close to the TX and the other close to the RX, that locally maximize $\rho$, and one position close to the middle of the TX-RX distance that minimizes it. Between the two local optima, the one closer to the TX is the global optimum for $h_t>h_r$, whereas it is the opposite for $h_t<h_r$. Moreover, the difference in magnitude between the two optima is relatively small for small values of $r_h$, but it becomes notable for relatively large values. This indicates that, from a practical point of view, in the former case the RIS can be placed either closer to the TX or RX with small difference in the resulting SNR, but in the latter case the SNR can notably differ and, hence, the optimal placement provided by the analytical model should be followed.
	
	\item As $y_s$ increases, the magnitude difference between the optima and the minimum substantially reduces and the location of the optima move closer to the middle of the TX-RX distance. Furthermore, there is a threshold $y_s$ value above which there is only one real value of $r_{1,h}$ that maximizes $\rho$ that is close to the middle of the TX-RX distance. This means that under such conditions the RIS can be practically placed close to the middle of the TX-RX distance regardless of whether $h_t>h_r$ or $h_r<h_t$.
	 
\end{itemize}

\subsubsection{$S_i\le S_s$ case}

\begin{itemize}
	\item Regardless of whether $h_t>h_r$ or $h_r<h_t$, there is only one optimum of $\rho$ with respect to $r_{1,h}$ that is located closer to the RX. In addition, a substantially higher $\rho$ is achieved by placing the RIS at the particular point rather than close to the TX or the middle of the TX-RX distance.
	
	\item $D_r$ should be notably smaller than $D_t$ so that \eqref{received_power_RIS larger than the footprint} holds. Under such a condition, $G_{r,n}$ is approximately constant throughout the illuminated RIS region and approximately equal to $G_r^{max}$. Equivalently, the RX antenna aperture needs to be sufficiently small so that the footprint of the highly directional departing beam from the RIS on the RX antenna is larger than the aperture of the latter.
\end{itemize}

Finally, we would like to note that the same analytical methodology  for extracting the optimal RIS placement in fixed-topology scenarios can be used for extracting the optimal placement also in mobile scenarios. In the latter case, due to fading that would likely arise as the user is moving, the analysis should consider the statistical average power effect of the fading process on the RIS-RX channel that would depend on the elevation of the RIS with respect to the ground.

	\section{Conclusions}\label{S:Conclusions}
	This work has been motivated by the need to answer the question of where an RIS that aids a highly-directional mmWave TX-RX link of fixed topology under blockage should be placed, so as to maximize the SNR performance. Based on this, we have firstly computed the end-to-end received power and SNR under an RIS of arbitrary size. Subsequently, we provided closed-form approximate expressions for the cases of the RIS being either much smaller or larger than the transmit beam footprint at the RIS plane. Finally, based on the resulting SNR expressions, we analytically derived the optimal horizontal RIS placement that maximizes the end-to-end SNR.
	
	The analytical outcomes have been validated by an extensive simulation campaign in various scenarios, which reveal that: i) when the transmission beam footprint at the RIS plane is much larger than the RIS size, the optimal RIS placement is either close to the TX, RX, or the middle of the TX-RX horizontal distance, depending on the system parameters; ii) when the footrpint is equal to or smaller than the RIS size, the optimal RIS placement is close to the RX. Such outcomes can be readily used by the system designer to properly deploy RISs in a way that the system performance is maximized. 
	
	\section*{Acknowledgements}
	The authors would like to cordially thank the associate editor and anonymous reviewers, whose comments and suggestions have led to a substantial improvement of this work.
	
	\section*{Appendices}
	\section*{Appendix A}
	\section*{Proof of Lemma~\ref{transmit_energy_main_lobe}}
	Based on \eqref{electric_field_radiation_pattern}, for the total normalized transmit power, which we denote by $P_{dish}^{tot}$, it holds that
	\begin{align}
		P_{dish}^{tot}=\int_{-\frac{\pi}{2}}^{\frac{\pi}{2}}E^2\left(\phi\right)d\phi.
		\end{align}
	As far as $\phi_0$ is concerned, which encompasses the main lobe, it can be computed by finding the points for which $\frac{\pi D_t \sin\left(\phi\right)}{\lambda}=3.83$ since $J_1\left(3.83\right)=0$. Hence, it holds that
	\begin{align}
		\phi_0=2\sin^{-1}\left(\frac{3.83\lambda}{\pi D_t}\right)=2\sin^{-1}\left(\frac{1.22\lambda}{D_t}\right).
	\end{align}
Consequently, the normalized transmit power that is corresponding to the main lobe, which we denote by $P_{dish}^{main\;lobe}$, is given by
\begin{align}
	P_{dish}^{main\;lobe}=\int_{-\frac{\phi_0}{2}}^{\frac{\phi_0}{2}}E^2\left(\phi\right)d\phi.
\end{align}
Let us denote the ratio of the normalized power inside the main lobe over the total normalized transmit power as a function of $\phi_{HPBW}$ by $\rho_{main\;lobe}\left(\phi_{HPBW}\right)$. Consequently, it holds that  $\rho_{main\;lobe}\left(\phi_{HPBW}\right)=P_{dish}^{main\;lobe}/P_{dish}^{tot}$ and for $\phi_{HPBW}<\frac{\pi}{12}$ it holds that  $\rho_{main\;lobe}\left(\phi_{HPBW}\right)>\rho_{main\;lobe}\left(\frac{\pi}{12}\right)$. $\frac{\pi}{12}$ corresponds to the 15$^{\circ}$ minimum HPBW value upper limit for which the transmission can be considered as pencil beam according to our assumptions. $\rho_{main\;lobe}\left(\phi_{HPBW}\right)>\rho_{main\;lobe}\left(\frac{\pi}{12}\right)$ holds due to the fact that as $\phi_{HPBW}$ decreases the main lobe becomes sharper, which means that more power is concentrated inside the main lobe compared with the $\phi_{HPBW}=\frac{\pi}{12}$ case. Hence, $\rho_{main\;lobe}\left(\phi_{HPBW}\right)$ increases.

As far as the value of the term $\frac{D_t}{\lambda}$ for which $\phi_{HPBW}=\frac{\pi}{12}$ is achieved, it holds that
\begin{small}
\begin{align}
	E^2\left(\frac{\pi}{24}\right)=0.5&\Rightarrow\left(\frac{2\lambda}{\pi D_t}\frac{J_1\left(\frac{\pi D_t \sin\left(\frac{\pi}{24}\right)}{\lambda}\right)}{\sin\left(\frac{\pi}{24}\right)}\right)^2=0.5\nonumber\\
	&\Rightarrow\frac{D_t}{\lambda}=3.94.
\end{align}
\end{small}
For $\frac{D_t}{\lambda}=3.94$ under which $\phi_{HPBW}=\frac{\pi}{12}$ is achieved, it holds that
\begin{small}
\begin{align}
	\rho_{main\;lobe}\left(\frac{\pi}{12}\right)&=\frac{\int_{-\sin^{-1}\left(\frac{1.22}{3.94}\right)}^{\sin^{-1}\left(\frac{1.22}{3.94}\right)}\left(\frac{2}{\pi 3.94}\frac{J_1\left(\pi 3.94 \sin\left(\phi\right)\right)}{\sin\left(\phi\right)}\right)^2d\phi}{\int_{-\frac{\pi}{2}}^{\frac{\pi}{2}}\left(\frac{2}{\pi 3.94}\frac{J_1\left(\pi 3.94 \sin\left(\phi\right)\right)}{\sin\left(\phi\right)}\right)^2d\phi}\nonumber\\
	&=0.97.
\end{align}
\end{small}
By taking into account that $\rho_{main\;lobe}\left(\phi_{HPBW}\right)>\rho_{main\;lobe}\left(\frac{\pi}{12}\right)$ for $\phi_{HPBW}<\frac{\pi}{12}=15^{\circ}$, the proof of Lemma~\ref{transmit_energy_main_lobe} is concluded.

	\section*{Appendix B}
	\section*{Proof of Lemma 2}
	
	By applying the law of sines in the ABC triangle, we obtain
	\begin{align}
		\frac{\sin\left(\frac{\phi_0}{2}\right)}{\alpha} = \frac{\sin\left(\overset{\wedge}{\text{C}} \right)}{r_1},
		\label{Eq:sine_laws}
	\end{align}
	where $\overset{\wedge}{\text{C}}$ denotes the angle of corner C and can be calculated~as 
	\begin{align}
		\overset{\wedge}{\text{C}} = \frac{\pi}{2} - \frac{\phi_0}{2} - \theta_i.
		\label{Eq:BCA_angle}
	\end{align}
	By substituting~\eqref{Eq:BCA_angle} into \eqref{Eq:sine_laws}, we obtain \eqref{Eq:alpha}.

	The eccentricity of the elliptical footprint can be evaluated as
	\begin{align}
		\epsilon=\frac{\sin\left(\theta_i\right)}{\sin\left(\frac{\pi}{2}-\frac{\phi_{0}}{2}\right)},
		\label{Eq:epsilon_s0}
	\end{align}
	or equivalently as in~\eqref{Eq:epsilon}. Finally, $\beta$ can be obtained as in~\eqref{Eq:beta}. This concludes the proof.

	\section*{Appendix C}
	\section*{Proof of Proposition~\ref{Proposition_1}}
	
	The incident electric field on the $n_{th}$ RU of the RIS illuminated area can be obtained~as
	\begin{align}
		\textbf{E}_{n}=E_{n} e^{-j\frac{2\pi}{\lambda}r_{1, n}}{\mathbf{n}}_o, \quad n=1,2,...,M,
		\label{Eq:Ein}
	\end{align}
	where $E_{n}$ is the amplitude of the incident wave  and ${\mathbf{n}}_o$ is a unitary vector that is perpendicular to the 2D plane that the electric field lies on \cite[Example 11-3]{balanis2012advanced}. It holds that
	\begin{equation}
	    \label{electric_field_amplitude}
		E_{n}=\sqrt{\frac{2\eta P_tG_{t,n}}{4\pi r_{1,n}^2}},
	\end{equation}where $\eta$ is the free-space impedance. 
    As a consequence, the power density at the $n_{th}$ RU of the RIS can be expressed~as 
	\begin{align}
		P_n = \frac{E_{n}^2}{2\eta}
	\end{align}
	or, with the aid of~\eqref{electric_field_amplitude},~as
	\begin{align}
		P_n = \frac{P_{t}G_{t,n}}{4 \pi r_{1,n}^2},
		\label{Eq:Pi_2}
	\end{align}
	where $\eta$ is the free-space impedance. Thus, the incident power at the $n_{th}$ RU can be evaluated~as
	\begin{align}
		P_{i,n} = P_n A_n,
		\label{Eq:Pin}
	\end{align}
	where $A_n$ stands for the effective aperture of the $n_{th}$ illuminated RU and can be obtained~as
	\begin{align}
		A_n = \frac{\lambda^2}{4\pi}G_{s}{\left(\theta_{i,n}\right)}.
		\label{Eq:A}
	\end{align}
	By substituting ~\eqref{Eq:Pi_2} and~\eqref{Eq:A} into~\eqref{Eq:Pin}, we obtain
	\begin{align}
		P_{i,n}=\left(\frac{\lambda}{4\pi}\right)^2
		\frac{P_{t}G_{t,n}G_{s}\left(\theta_{i,n}\right)}{r_{1,n}^2}.
		\label{Eq:Pin_2}
	\end{align}
	As a result and due to the energy conservation law, the reflected power density by the $n_{th}$ RU, which is captured by the RX antenna, can be expressed~as
	\begin{align}
		P_{r,n} =  \frac{P_{i,n} \Gamma^{2} G_{s}\left(\theta_{r,n}\right)}{4\pi r_{2,n}^2}.
		\label{Eq:Prn}
	\end{align}
	By substituting~\eqref{Eq:Pin_2} into~\eqref{Eq:Prn}, we obtain
	\begin{align}
		P_{r,n} =\frac{\lambda^2}{\left(4 \pi\right)^3}\frac{P_{t}\Gamma^2G_{t,n}G_{s}{\left(\theta_{i,n}\right)}G_{s}{\left(\theta_{r,n}\right)}}{r_{1,n}^2 r_{2,n}^2}.
		\label{Eq:Prn2}
	\end{align}
The power captured by the receiver from the $n_{th}$ RU is given by
\begin{align}
	P_{R,n}&=P_{r,n}\frac{\lambda^2}{4\pi}G_{r,n}\nonumber\\
	&=\left(\frac{\lambda}{4 \pi}\right)^4\frac{P_{t}\Gamma^2G_{t,n}G_{r,n}G_{s}{\left(\theta_{i,n}\right)}G_{s}{\left(\theta_{r,n}\right)}}{r_{1,n}^2 r_{2,n}^2}.
\end{align}
	Moreover, the corresponding electric field observed at the receiver from the $n_{th}$ RU can be evaluated~as
	\begin{equation}
		\mathbf{E}_{R,n}=E_{R,n} e^{-j\left(\theta_{n}+\frac{2\pi\left(r_{1,n}+r_{2,n}\right)}{\lambda}\right)} {\mathbf{a}_o},
		\label{Eq:E_rn}
	\end{equation}
	where $\mathbf{a}_o$ is a unitary vector perpendicular to the 2D plane that the reflected electric field lies on and $\theta_n$ is adjustable the phase response of the $n_{th}$ RU. Additionally, $E_{R,n}$ can be computed~as  
	\begin{align}
		E_{R,n}&=\sqrt{2 \eta P_{R,n}},
	\end{align}
	which, by employing~\eqref{Eq:Prn2}, can be rewritten~as
	\begin{align}
		E_{R,n}&=\sqrt{2\eta\left(\frac{\lambda}{4 \pi}\right)^4\frac{P_{t}\Gamma^2G_{t,n}G_{r,n}G_{s}{\left(\theta_{i,n}\right)}G_{s}{\left(\theta_{r,n}\right)}}{r_{1,n}^2 r_{2,n}^2}}.
		\label{Eq:E_rn3}
	\end{align}
	
	From~\eqref{Eq:E_rn}, the aggregated electric field at the RX can be written~as
	\begin{align}
		\mathbf{E}_{R} = \sum_{n=1}^{M} \mathbf{E}_{R,n}.
	\end{align}
	
	Consequently, by employing~\eqref{Eq:E_rn} and~\eqref{Eq:E_rn3} it holds that
	\begin{align}
		\mathbf{E}_{R} &= \left(\frac{\lambda}{4\pi}\right)^2\sqrt{2\eta P_t\Gamma}\sum_{n=1}^{M}\sqrt{\frac{G_{t,n}G_{r,n}G_{s}{\left(\theta_{i,n}\right)}G_{s}{\left(\theta_{r,n}\right)}}{r_{1,n}^2 r_{2,n}^2}}
		\nonumber\\&\times e^{-j\left(\theta_{n}+\frac{2\pi\left(r_{1,n}+r_{2,r}\right)}{\lambda}\right)}{\mathbf{a}_o},
		\label{Eq:Er_final}
	\end{align}

	Hence, at the RX the received power can be obtained~as
	\begin{align}
		P_R = \frac{\left|\mathbf{E}_{R}\right|^2}{2\eta},
	\end{align}
	which, with the aid of~\eqref{Eq:Er_final}, can be rewritten~as
	\begin{small}
	\begin{align}
		P_R &= \left(\frac{\lambda}{4\pi}\right)^4P_t\Gamma^2\nonumber\\
		&\times\left|\sum_{n=1}^{M}\sqrt{\frac{G_{t,n}G_{r,n}G_{s}{\left(\theta_{i,n}\right)}G_{s}{\left(\theta_{r,n}\right)}}{r_{1,n}^2 r_{2,n}^2}}e^{-j\left(\theta_{n}+\frac{2\pi\left(r_{1,n}+r_{2,n}\right)}{\lambda}\right)}\right|^2.
		\label{received_power_RIS}
	\end{align}
\end{small}
	By assuming that the optimal phase shift is induced by each RU so as to maximize $P_R$, i.e.,
	\begin{align}
		\theta_n = -\frac{2\pi\left(r_{1,n}+r_{2,n}\right)}{\lambda},
	\end{align}
	the received power can be rewritten~as
	\begin{small}
		\begin{align}
			P_R &= \left(\frac{\lambda}{4\pi}\right)^4P_t\Gamma^2
			\left|\sum_{n=1}^{M}\sqrt{\frac{G_{t,n}G_{r,n}G_{s}{\left(\theta_{i,n}\right)}G_{s}{\left(\theta_{r,n}\right)}}{r_{1,n}^2 r_{2,n}^2}}\right|^2,
			\label{received_power_RIS}
		\end{align}
	\end{small}
	which concludes the~proof.

	\section*{Appendix D}
	\section*{Proof of Lemma~\ref{Lemma_energy_step_function_HPBW}}
	
	Let us denote the ratio of the normalized power included within a step function in the interval $\left[-\frac{\phi_{HPBW}}{2},\frac{\phi_{HPBW}}{2}\right]$ over the normalized power included within the FNBW by $\kappa\left(\phi_{HPBW}\right)$. It holds that
	\begin{small}
	\begin{align}
		\kappa\left(\phi_{HPBW}\right)=\frac{\phi_{HPBW}}{P_{dish}^{main\;lobe}}=\frac{\phi_{HPBW}}{\int_{-\frac{\phi_0}{2}}^{\frac{\phi_0}{2}}\left(\frac{2\lambda}{\pi D_t}\frac{J_1\left(\frac{\pi D_t \sin\left(\phi\right)}{\lambda}\right)}{\sin\left(\phi\right)}\right)^2d\phi}.
	\end{align}
\end{small}
For the minimum HPBW of value of 15$^{\circ}$ required for the transmit beam to be considered pencil beam, it holds that $\kappa\left(\frac{\pi}{12}\right)=0.97$. Moreover,  
$\kappa\left(\phi_{HPBW}\right)$ is a monotonically increasing function as $\phi_{HPBW}\downarrow$ since the smaller the beamwidth is the sharper the main lobe becomes. Consequently, it can be more accurately approximated by a step function with magnitude equal to $G_t^{max}$ in the interval $\left[-\frac{\phi_{HPBW}}{2},\frac{\phi_{HPBW}}{2}\right]$. 

In addition, by denoting the ratio of the normalized power included within the step function over the total normalized power that impinges on the RIS by $\mu\left(\phi_{HPBW}\right)$, it holds that

\begin{small}
	\begin{align}
		\mu\left(\phi_{HPBW}\right)=\frac{\phi_{HPBW}}{P_{dish}^{tot}}=\frac{\phi_{HPBW}}{\int_{-\frac{\pi}{2}}^{\frac{\pi}{2}}\left(\frac{2\lambda}{\pi D_t}\frac{J_1\left(\frac{\pi D_t \sin\left(\phi\right)}{\lambda}\right)}{\sin\left(\phi\right)}\right)^2d\phi}.
	\end{align}
\end{small}
For the minimum HPBW of value of 15$^{\circ}$ it holds that $\mu\left(\frac{\pi}{12}\right)=0.94$. Moreover, for the same reason as in the 	$\kappa\left(\phi_{HPBW}\right)$ case
$\mu\left(\phi_{HPBW}\right)$ is a monotonically increasing function as $\phi_{HPBW}\downarrow$. 
	
\section*{Appendix E}
\section*{Proof of Proposition~\ref{Proposition_Optimal_RIS_placement_small}}	
The proof begins by plugging \eqref{received_power_RIS_Proposition_2} into \eqref{end_to_end_SNR} and rewriting $\rho$ as
\begin{align}
    \label{SNR_small_optimal_point_proof}
	\rho=\left(\frac{\lambda}{4\pi}\right)^4
	\frac{16P_t\Gamma^2\left(S_s\right)^2G_{t}^{max}G_{r}^{max}}{d_x^2d_y^2N_0}	\frac{F\left(r_{1,h}\right)}{G\left(r_{1,h}\right)},
\end{align}
	where
	\begin{align}
		F\left(r_{1,h}\right)&{=}\cos\left(\tan^{-1}\left(\frac{\sqrt{r_{{1},{h}}^2+\left(h_s-h_{t}\right)^2}}{y_s}\right)\right)
		\nonumber\\&\times
		\cos\left(\tan^{-1}\left(\frac{\sqrt{\left(r_{{1},{h}}-r_h\right)^2+\left(h_s-h_{r}\right)^2}}{y_s}\right)\right)
		\label{Small_RIS_term_1}
	\end{align}
	and
	\begin{align}
		G\left(r_{1,h}\right)&{=}\left(r_{{1},{h}}^2+y_s^2+\left(h_{s}-h_{t}\right)^2\right)\nonumber\\
		&\times\left(\left(r_{h}-r_{{1,h}}\right)^2+y_s^2+\left(h_{s}-h_{r}\right)^2\right).
		\label{Small_RIS_term_2}
	\end{align}
	
	From~\eqref{SNR_small_optimal_point_proof},  we observe that  the end-to-end SNR depends on 
	$r_{1,h}$ through the ratio $\frac{F\left(r_{1,h}\right)}{G\left(r_{1,h}\right)}$. Hence, the optimum $r_{1,h}$ that maximizes $\rho$ can be obtained by evaluating the roots of the first derivative of $\frac{F\left(r_{1,h}\right)}{G\left(r_{1,h}\right)}$ with respect to $r_{1,h}$. The first derivative of $\frac{F\left(r_{1,h}\right)}{G\left(r_{1,h}\right)}$ can be obtained as in~\eqref{1st_Derivate_of_SNR_small}, given at the top of the following page. Consequently, $r_{1,h}^{*}$ is obtained as one of the solutions of $a^{\left(1\right)} r_{1,h}^3 + b^{\left(1\right)} r_{1,h}^2 + c^{\left(1\right)} r_{1,h} + d^{\left(1\right)} = 0$.  
	\begin{figure*}
		\begin{small}  
		\begin{align}
			\frac{d\left(\frac{F\left(r_{1,h}\right)}{G\left(r_{1,h}\right)}\right)}{dr_{1,h}}&=-\left(6r_{1,h}^3-9r_hr_{1,h}^2+3\left(2y_s^2+r_h^2+\left(h_s-h_t\right)^2+\left(h_s-h_r\right)^2\right)r_{1,h}-3r_h\left(y_s^2+\left(h_s-h_t\right)^2\right)\right)\nonumber\\
			&\times \left(\frac{\left(y_s^2+r_{1,h}^2+\left(h_s-h_t\right)^2\right)\left(y_s^2+\left(r_h-r_{1,h}\right)^2+\left(h_s-h_r\right)^2\right)}{\left(h_s-h_t\right)^2\left(h_s-h_r\right)^2}\right)^{-\frac{1}{2}}\nonumber\\
			&\times\left(\left(y_s^2+r_{1,h}^2+\left(h_s-h_t\right)^2\right)\left(y_s^2+\left(r_h-r_{1,h}\right)^2+\left(h_s-h_r\right)^2\right)\right)^{-2}.
			\label{1st_Derivate_of_SNR_small}
		\end{align}
	    \end{small}
		\hrulefill
	\end{figure*}
	
	\section*{Appendix F}
	\section*{Proof of Proposition~\ref{Proposition_Optimal_RIS_placement_large}}
	The proof begins by plugging \eqref{Eq:P_R} into \eqref{end_to_end_SNR} and rewriting $\rho$ as
\begin{align}
	\label{SNR_large_optimal_point_proof}
	\rho&\approx\left(\frac{\lambda}{4\pi}\right)^4
	\frac{16P_t\Gamma^2G_{t}^{max}G_{r}^{max}\pi^2\sin^4\left(\frac{\phi_{HPBW}}{2}\right)}{d_x^2d_y^2N_0}\nonumber\\	
	&\times F\left(r_{1,h}\right)H\left(r_{1,h}\right),\end{align}where
\begin{small}
\begin{align}
	\label{term_RIS_large_first_derivative}
	H\left(r_{1,h}\right)=\frac{r_{{1},{h}}^2+y_s^2+\left(h_{s}-h_{t}\right)^2}{\left(r_{h}-r_{{1,h}}\right)^2+y_s^2+\left(h_{s}-h_{r}\right)^2}\frac{1-\frac{\sin^2\left(\theta_i\right)}{\cos^2\left(\frac{\phi_{HPBW}}{2}\right)}}{\cos^4\left(\frac{\phi_{HPBW}}{2}+\theta_i\right)}.
	\end{align}
\end{small}
 Hence, the optimum $r_{1,h}$ that maximizes $\rho$ can be obtained by evaluating the roots of the first derivative of $F\left(r_{1,h}\right)H\left(r_{1,h}\right)$ with respect to $r_{1,h}$. Before computing the corresponding derivative, we simplify things by taking into that for pencil-beam transmissions it holds that $\phi_{HPBW}<<1$. Hence, considering that $\cos\left(\frac{\phi_{HPBW}}{2}\right)\approx 1$ and $\cos\left(\frac{\phi_{HPBW}}{2}+\theta_i\right)\approx \cos\left(\theta_i\right)$, $H\left(r_{1,h}\right)$ can be approximated as 
 \begin{align}
 	H\left(r_{1,h}\right)\approx\frac{r_{{1},{h}}^2+y_s^2+\left(h_{s}-h_{t}\right)^2}{\left(r_{h}-r_{{1,h}}\right)^2+y_s^2+\left(h_{s}-h_{r}\right)^2}\frac{1}{\cos^2\left(\theta_i\right)}.
 \end{align}
As a result, the first derivative of $F\left(r_{1,h}\right)H\left(r_{1,h}\right)$ can be approximated by \eqref{first_derivative_RIS_large_approximation} given at the top of the next page.
\begin{figure*}
	\begin{small}
\begin{align}
	\label{first_derivative_RIS_large_approximation}
	\frac{d\left(F\left(r_{1,h}\right)H\left(r_{1,h}\right)\right)}{dr_{1,h}}\approx-\frac{3\sqrt{(r_{1,h}^2+y_s^2+\left(h_s-h_t\right)^2}\left(r_hr_{1,h}^2+\left(\left(h_s-h_t\right)^2-r_h^2-\left(h_s-h_r\right)^2\right)r_{1,h}-r_h\left(y_s^2+\left(h_s-h_t\right)^2\right)\right)}{\left(\left(r_{h}-r_{{1,h}}\right)^2+y_s^2+\left(h_{s}-h_{r}\right)^2\right)^{\frac{5}{2}}}.
\end{align}
\end{small}
\hrulefill
\end{figure*}
Consequently, $r_{1,h}$ is obtained as one of the two solutions of $r_hr_{1,h}^2+\left(\left(h_s-h_t\right)^2-r_h^2-\left(h_s-h_r\right)^2\right)r_{1,h}-r_h\left(y_s^2+\left(h_s-h_t\right)^2\right)=0$. As it is shown in Section~\ref{S:Res}, between its two real roots the one that maximizes $\rho$ is the one closer to RX, given by \eqref{solution_RIS_large}. 
\begin{small}
\begin{align}	
	\end{align}
\end{small}
	
	\bibliographystyle{IEEEtran}
	\bibliography{IEEEabrv,references}
	
	\begin{IEEEbiography} [{\includegraphics[width=1in,height=1.25in,clip,keepaspectratio]{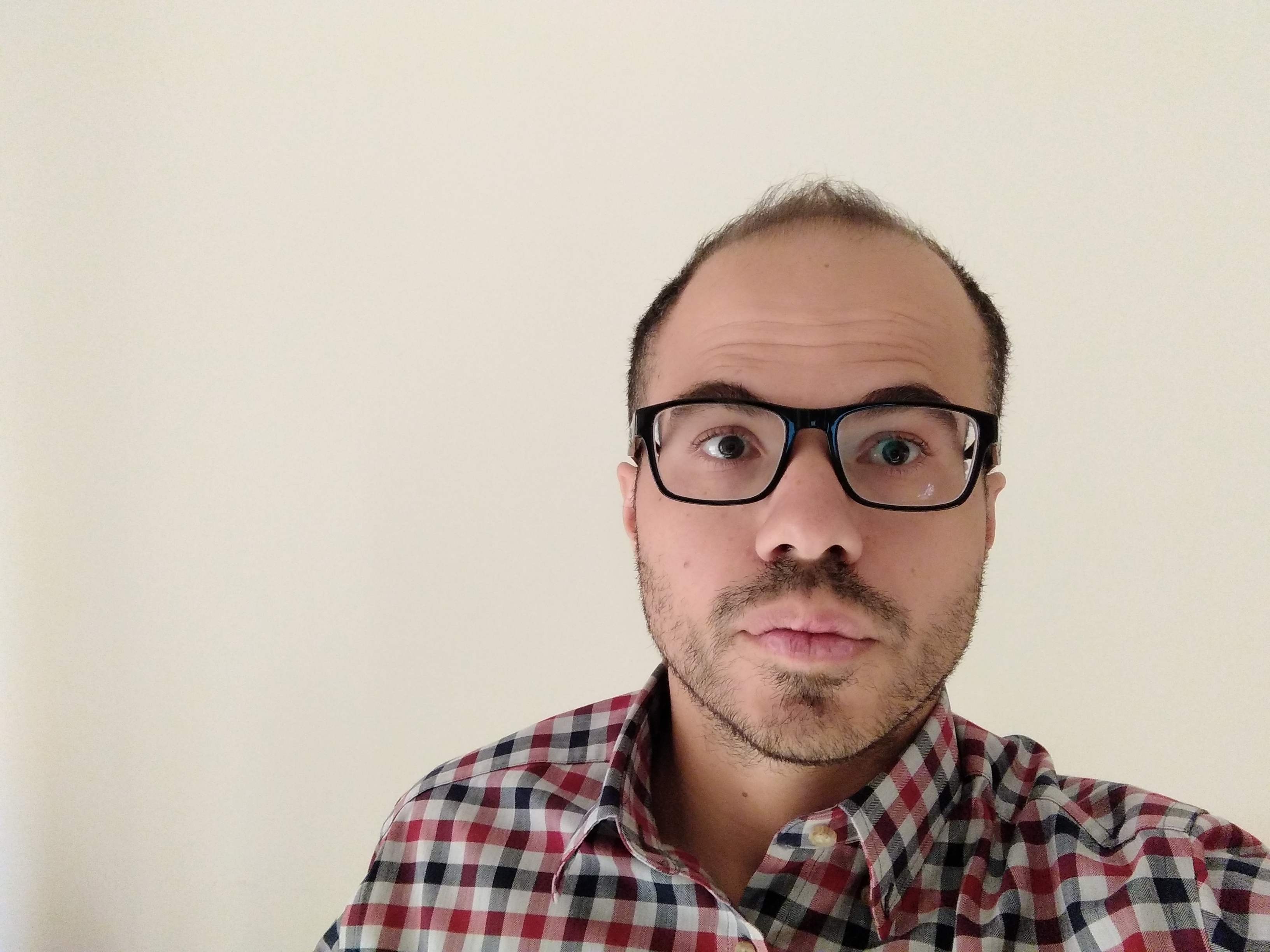}}] 
		{Konstantinos Ntontin} (S’12-M’14) is currently a research associate of the SIGCOM Research Group at SnT, University of Luxembourg. In the past, he held research associate positions at the Electronic Engineering and Telecommunications department of the University of Barcelona and at the Informatics and Telecommunications department of the University of Athens. In addition, he held an internship position at Ericsson Eurolab Gmbh, Germany. He received the Diploma in Electrical and Computer Engineering in 2006, the M. Sc. Degree in Wireless Systems in 2009, and the Ph. D. degree in 2015 from the University of Patras, Greece, the Royal Institute of Technology (KTH), Sweden, and the Technical University of Catalonia (UPC), Spain, respectively. His research interests are related to the physical layer of wireless telecommunications with focus on performance analysis in fading channels, MIMO systems, array beamforming, transceiver design, and stochastic modeling of wireless channels.	
	\end{IEEEbiography}

	\begin{IEEEbiography}[{\includegraphics[width=1in,height=1.25in,clip,keepaspectratio]{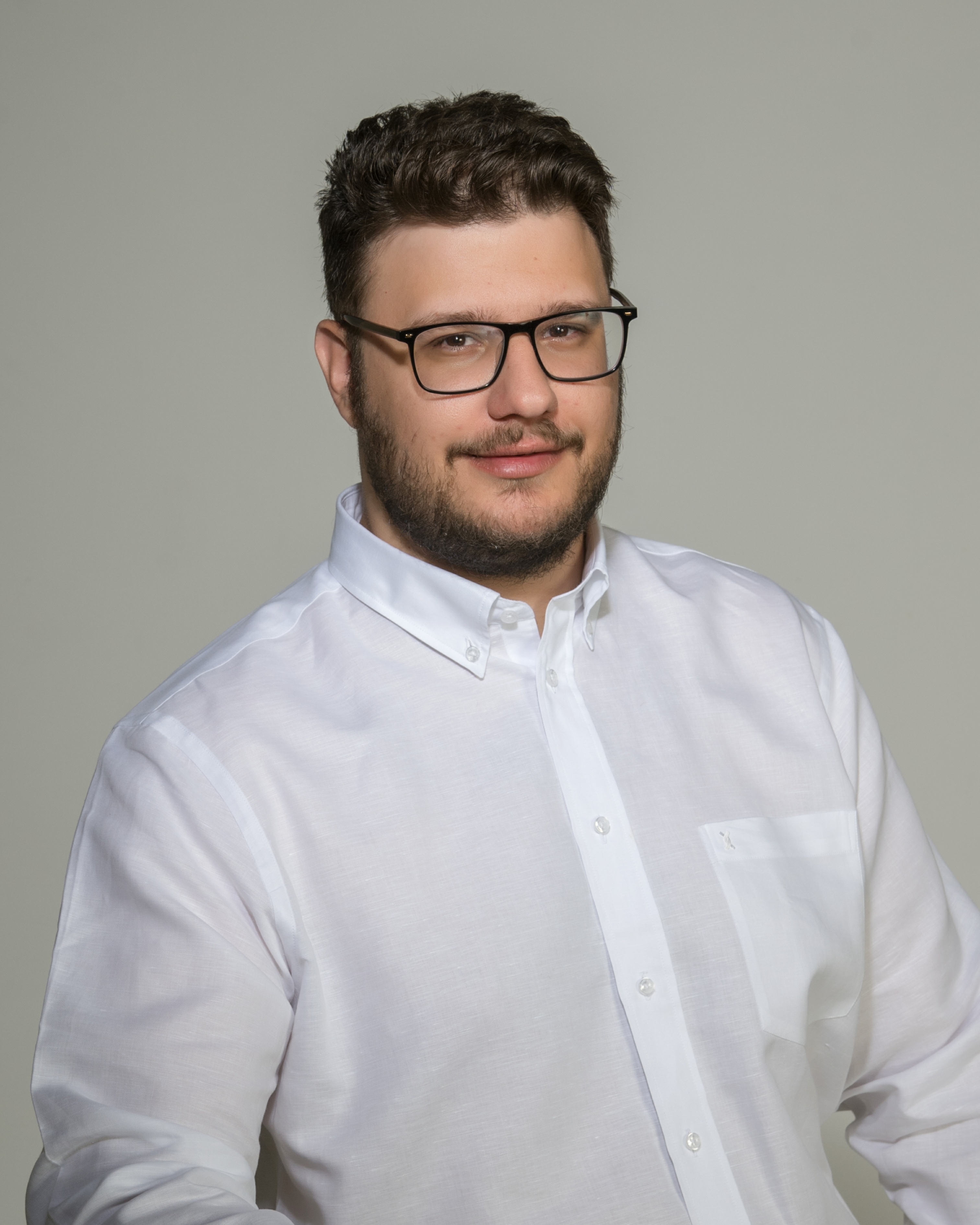}}]{Alexandros-Apostolos A. Boulogeorgos} (S'11, M'16, SM'19) was born in Trikala, Greece in 
		1988. He received the Electrical and Computer Engineering (ECE) diploma degree and Ph.D. degree in Wireless Communications from the Aristotle University of Thessaloniki (AUTh) in 2012 and 2016, respectively. 
		
		From November 2012, he has been a member of the wireless communications system group of AUTh, working as a research assistant/project engineer in various telecommunication and networks projects. During 2017, he joined the information technologies institute, while from November 2017, he has joined the Department of Digital Systems, University of Piraeus, where he conducts research  in the area of wireless communications. Moreover, from October 2012 until September 2016, he was a teaching assistant at the department of ECE of AUTh, whereas, from February 2017, he serves as an adjunct lecturer at the Department of Informatics and Telecommunications Engineering of the University of Western Macedonia and as an visiting lecturer at the Department of Computer Science and Biomedical Informatics of the University of Thessaly.  
		
		Dr. Boulogeorgos  has authored and co-authored more than 65 technical papers, which were published in scientific journals and presented at prestigious international conferences. Furthermore, he has submitted two (one national and one European) patents. Likewise, he has been involved as member of Technical Program Committees in several IEEE and non-IEEE conferences and served as a reviewer in various IEEE journals and conferences. 
		Dr. Boulogeorgos was awarded with the ``Distinction Scholarship Award'' of the Research Committee of AUTh for the year 2014 and was recognized as an exemplary reviewer for IEEE Communication Letters for 2016 (top $3\%$ of reviewers). Moreover, he was named a top peer reviewer (top $1\%$ of reviewers) in Cross-Field and Computer Science in the Global Peer Review Awards 2019, which was presented by the Web of Science and Publons.  His current research interests spans in the area of wireless communications and networks with emphasis in high frequency communications, optical wireless communications and communications for biomedical applications.
		He is a Senior Member of the IEEE and a member of the Technical Chamber of Greece. He is currently an  Editor for IEEE Communications Letters, and an Associate Editor for the Frontier In Communications And Networks.
	\end{IEEEbiography}
	
	\begin{IEEEbiography} [{\includegraphics[width=1in,height=1.25in,clip,keepaspectratio]{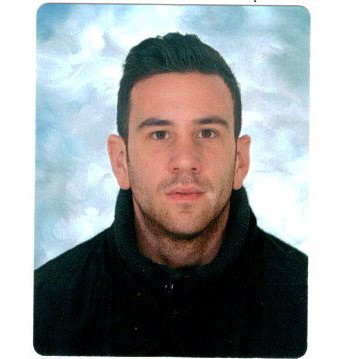}}] 
		{Dimitrios G. Selimis} (S'20) received his Diploma degree in Electrical and Computer Engineering from the University of Patras, Greece, in 2018. He received his master degree in Modern Wireless Communications from University of Peloponnese in 2019. His master thesis focused on the performance analysis of Spatial Modulation-MIMO systems for several fading scenarios. He is currently a PhD candidate at University of Peloponnese in collaboration with the National Centre for Scientific Research-``Demokritos''. His current research interests are related to the physical layer of wireless communications with focus on statistical modeling of wireless channels.	
	\end{IEEEbiography}
	
	\begin{IEEEbiography} [{\includegraphics[width=1in,height=1.25in,clip,keepaspectratio]{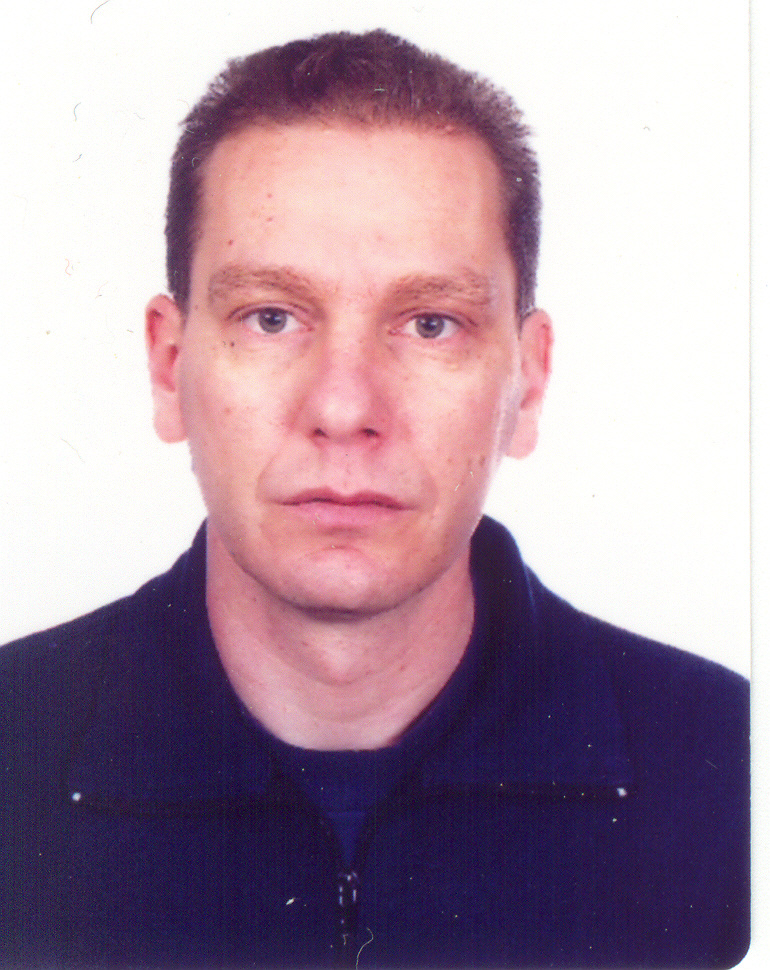}}] {Fotis I. Lazarakis} received his diploma in Physics in 1990, from Department of Physics, Aristotle University of Thessaloniki, Greece, and his Ph.D in Mobile Communications, in 1997, from Department of Physics, National and Kapodistrian University of Athens, Greece, holding at the same time a scholarship from National Center for Scientific Research "Demokritos" (NCSRD), Institute of Informatics and Telecommunications (IIT). From 1999 to 2002 he was with Telecommunications Laboratory, National Technical University of Athens, as a senior research associate. In 2003 he joined NCSRD, Institute of Informatics and Telecommunications as a Researcher and since 2013 is a Research Director. He has been involved in a number of national and international projects, acting as a Project Manager to several of those. His research interests include WLANs, 5G and beyond, propagation models and measurements, fading channel characteristics and capacity, diversity techniques, MIMO antennas and systems, radio resource management and performance evaluation of mobile/wireless networks. Dr. Lazarakis has authored or co-authored more than 100 journal and conference papers and he is co-owner of a patent.
		
	\end{IEEEbiography}

	\begin{IEEEbiography} [{\includegraphics[width=1in,height=1.25in,clip,keepaspectratio]{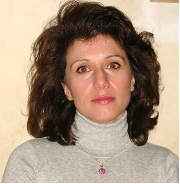}}] {Angeliki Alexiou} is a professor at the department of Digital Systems, ICT School, University of Piraeus. She received the Diploma in Electrical and Computer Engineering from the National Technical University of Athens in 1994 and the PhD in Electrical Engineering from Imperial College of Science, Technology and Medicine, University of London in 2000. Since May 2009 she has been a faculty member at the Department of Digital Systems, where she conducts research and teaches undergraduate and postgraduate courses in the area of Broadband Communications and Advanced Wireless Technologies. Prior to this appointment she was with Bell Laboratories, Wireless Research, Lucent Technologies, (later Alcatel-Lucent, now NOKIA), in Swindon, UK, first as a member of technical staff (January 1999-February 2006) and later as a Technical Manager (March 2006-April 2009). Professor Alexiou is a co-recipient of Bell Labs President’s Gold Award in 2002 for contributions to Bell Labs Layered Space-Time (BLAST) project and the Central Bell Labs Teamwork Award in 2004 for role model teamwork and technical achievements in the IST FITNESS project. Professor Alexiou is the Chair of the Working Group on Radio Communication Technologies and of the Working Group on High Frequencies Radio Technologies of the Wireless World Research Forum. She is a member of the IEEE and the Technical Chamber of Greece. Her current research interests include radio interface for 5G systems and beyond, MIMO and high frequencies (mmWave and THz wireless) technologies, cooperation, coordination and efficient resource management for Ultra Dense wireless networks and machine-to-machine communications, `cell-less' architectures based on virtualization and extreme resources sharing and machine learning for wireless systems. She is the project coordinator of the H2020 TERRANOVA project (ict-terranova.eu) and the technical manager of H2020 ARIADNE project (ict-ariadne.eu).
	\end{IEEEbiography}
	
	\begin{IEEEbiography} [{\includegraphics[width=1in,height=1.25in,clip,keepaspectratio]{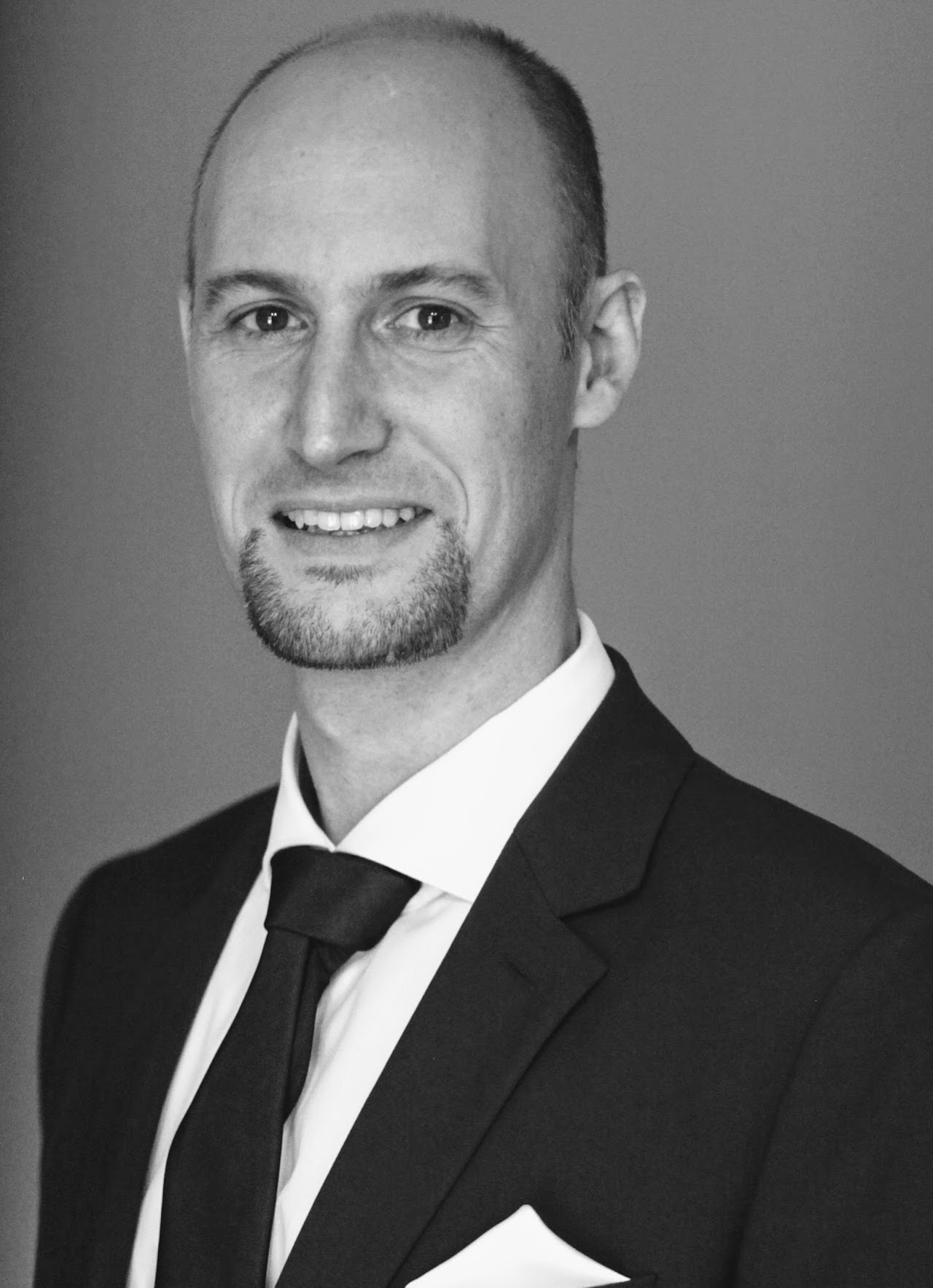}}] {Symeon Chatzinotas }
		(S’06–M’09–SM’13) is currently Full Professor / Chief Scientist I and Co-Head of the SIGCOM Research Group at SnT, University of Luxembourg. In the past, he has been a Visiting Professor at the University of Parma, Italy and he was involved in numerous Research and Development projects for the National Center for Scientific Research Demokritos, the Center of Research and Technology Hellas and the Center of Communication Systems Research, University of Surrey. He received the M.Eng. degree in telecommunications from the Aristotle University of Thessaloniki, Thessaloniki, Greece, in 2003, and the M.Sc. and Ph.D. degrees in electronic engineering from the University of Surrey, Surrey, U.K., in 2006 and 2009, respectively. He was a co-recipient of the 2014 IEEE Distinguished Contributions to Satellite Communications Award,  the CROWNCOM 2015 Best Paper Award and the 2018 EURASIC JWCN Best Paper Award. He has (co-)authored more than 400 technical papers in refereed international journals, conferences and scientific books. He is currently in the editorial board of the IEEE Open Journal of Vehicular Technology and the International Journal of Satellite Communications and Networking. 
		
	\end{IEEEbiography}

\end{document}